\documentclass[pdflatex,sn-nature]{sn-jnl}%

\usepackage{graphicx}%
\usepackage{subcaption}
\usepackage{multirow}%
\usepackage{amsmath,amssymb,amsfonts}%
\usepackage{amsthm}%
\usepackage[title]{appendix}%
\usepackage[table,xcdraw]{xcolor}
\usepackage{textcomp}%
\usepackage{manyfoot}%
\usepackage{booktabs}%
\usepackage{algorithm}%
\usepackage{algorithmicx}%
\usepackage{algpseudocode}%
\usepackage{listings}%
\usepackage{lineno}
\usepackage{url}
\usepackage{caption}    %
\usepackage{tocloft}    %
\usepackage{hyperref}   %
\usepackage[version=4]{mhchem}
\usepackage[T1]{fontenc}
\usepackage{tablefootnote}

\theoremstyle{thmstyleone}%

\theoremstyle{thmstyletwo}%

\theoremstyle{thmstylethree}%

\newcounter{supptable}
\newcounter{suppfigure}

\renewcommand{\thesupptable}{\arabic{supptable}}
\renewcommand{\thesuppfigure}{\arabic{suppfigure}}

\newenvironment{supptable}[1][htbp]{
   ~\refstepcounter{supptable}
    \renewcommand{\thetable}{\thesupptable}
    \begin{table}[#1]
    \centering
    \def\supptablecaption##1##2{%
        \caption*{\textbf{Supplementary Table~\thesupptable. ##1.}%
        \ifx##2\empty\else\ ##2\fi}%
        \addcontentsline{toc}{subsection}{Supplementary Table~\thesupptable. ##1}%
        \phantomcaption
    }
}{%
    \end{table}
}

\newenvironment{suppfigure}[1][htbp]{
   ~\refstepcounter{suppfigure}
   \renewcommand{\thefigure}{\thesuppfigure}
    \begin{figure}[#1]
    \centering
    \def\@captype{figure}
    \def\suppfigurecaption##1##2{%
        \caption*{\textbf{Supplementary Figure~\thesuppfigure. ##1.}
        \ifx##2\undefined\else~##2\fi}%
        \addcontentsline{toc}{subsection}{Supplementary Figure~\thesuppfigure. ##1}%
        \phantomcaption
    }
}{%
    \end{figure}
}

\newcounter{suppnotecounter}
\newcommand{\suppnote}[1]{%
   ~\refstepcounter{suppnotecounter}%
    \subsection*{Supplementary Note \arabic{suppnotecounter}. #1}%
    \addcontentsline{toc}{subsection}{Supplementary Note \arabic{suppnotecounter}. #1}%
}

\makeatletter
\renewcommand{\p@subfigure}{\thefigure}          %
\makeatother
\begin{document}

\title{NMR-Solver: Automated Structure Elucidation via Large-Scale Spectral Matching and Physics-Guided Fragment Optimization}

\author[1,2]{\fnm{Yongqi} \sur{Jin}}\email{yongqijin@stu.pku.edu.cn}
\author[2,3]{\fnm{Jun-Jie} \sur{Wang}}\email{junjie-wang@pku.edu.cn}
\author[2,4]{\fnm{Fanjie} \sur{Xu}}\email{xufanjie@stu.xmu.edu.cn}
\author[2]{\fnm{Xiaohong} \sur{Ji}}\email{jixh@dp.tech}
\author[2]{\fnm{Zhifeng} \sur{Gao}}\email{gaozf@dp.tech}
\author[2,5]{\fnm{Linfeng} \sur{Zhang}}\email{zhanglf@dp.tech}

\author*[2]{\fnm{Guolin} \sur{Ke}}\email{kegl@dp.tech}
\author*[3,5]{\fnm{Rong} \sur{Zhu}}\email{rongzhu@pku.edu.cn}
\author*[1,5,6]{\fnm{Weinan} \sur{E}}\email{weinan@math.pku.edu.cn}

\affil[1]{\orgdiv{School of Mathematical Sciences}, \orgname{Peking University}, \orgaddress{\city{Beijing}, \postcode{100871}, \country{China}}}
\affil[2]{\orgname{DP Technology}, \orgaddress{\city{Beijing}, \postcode{100080}, \country{China}}}
\affil[3]{\orgdiv{College of Chemistry and Molecular Engineering}, \orgname{Peking University}, \orgaddress{\city{Beijing}, \postcode{100871}, \country{China}}}
\affil[4]{\orgdiv{College of Chemistry and Chemical Engineering}, \orgname{Xiamen University}, \orgaddress{\city{Xiamen}, \postcode{361005}, \country{China}}}
\affil[5]{\orgname{AI for Science Institute}, \orgaddress{\city{Beijing}, \postcode{100080}, \country{China}}}
\affil[6]{\orgname{Center for Machine Learning Research, Peking University}, \orgaddress{\city{Beijing}, \postcode{100871}, \country{China}}}

\abstract{
Nuclear Magnetic Resonance (NMR) spectroscopy is one of the most powerful and widely used tools for molecular structure elucidation in organic chemistry.
However, the interpretation of NMR spectra to determine unknown molecular structures remains a labor-intensive and expertise-dependent process, particularly for complex or novel compounds.
Although recent methods have been proposed for molecular structure elucidation, they often underperform in real-world applications due to inherent algorithmic limitations and limited high-quality data.
Here, we present \textbf{NMR-Solver}, a practical and interpretable framework for the automated determination of small organic molecule structures from $^1$H and $^{13}$C NMR spectra.
Our method introduces an automated framework for molecular structure elucidation, integrating large-scale spectral matching with physics-guided fragment-based optimization that exploits atomic-level structure–spectrum relationships in NMR.
We evaluate NMR-Solver on simulated benchmarks, curated experimental data from the literature, and real-world experiments, demonstrating its strong generalization, robustness, and practical utility in challenging, real-life scenarios.
NMR-Solver unifies computational NMR analysis, deep learning, and interpretable chemical reasoning into a coherent system. By incorporating the physical principles of NMR into molecular optimization, it enables scalable, automated, and chemically meaningful molecular identification, establishing a generalizable paradigm for solving inverse problems in molecular science.
}


\maketitle

\section{Introduction} \label{sec:intro}

Nuclear Magnetic Resonance (NMR) spectroscopy stands as the most informative and widely used technique for molecular structure elucidation in organic chemistry~\cite{clayden2012organic}.
Unlike mass spectrometry or infrared spectroscopy, which primarily yield fragmentation patterns or functional group, NMR provides atomic-level insights into molecular connectivity, stereochemistry, and spatial arrangement~\cite{skoog2019textbook}.
These characteristics make NMR indispensable for the structural characterization of both known and novel small organic molecules, delivering a comprehensive view of the complete molecular architecture.

Despite its analytical power, the interpretation of NMR spectra is a time-consuming and expertise-dependent process. 
Although computational techniques~\cite{elyashberg2008computer,ermanis2017doubling,howarth2020dp4,marcarino2020nmr} and simulation software~\cite{ACDLabs, Mestre} have advanced and assist in validation and checking, manual interpretation remains the standard practice in many laboratories.
This reliance on human expertise not only limits throughput but also introduces variability and potential errors, particularly for complex or unknown compounds. Compounding this challenge is the vastness of chemical space—estimated to encompass over $10^{60}$ plausible organic molecules~\cite{ruddigkeit2012enumeration}—which makes structure elucidation akin to finding a needle in a cosmic haystack.
With the growing adoption of high-throughput screening and automated synthesis platforms~\cite{trobe2018molecular}, there is an increasing need for rapid, accurate, and automated NMR-based structure elucidation to keep pace with the accelerated rate of molecular discovery.

Recent advances in deep learning have significantly improved the accuracy and efficiency of predicting chemical shifts from known molecular structures—a key component of forward NMR simulation~\cite{jonas2019FCG, han2022SGNN, zou2023DeltaNet}. 
Both 2D graph-based models and 3D conformation-aware architectures now achieve accurate predictions comparable to density functional theory (DFT)~\cite{wolinski1990giao} calculations, with orders of magnitude faster inference, exemplified by GT-NMR~\cite{chen2024gtnmr} and NMRNet~\cite{xu2025toward}, respectively.
These capabilities have enabled large-scale spectral matching and provide a foundation for addressing the more challenging inverse problem: determining molecular structures from experimental NMR spectra.

In parallel with advances in forward modeling, several approaches have been proposed to address this inverse problem, broadly falling into two categories—AI generative models and traditional approaches.
Among the recent generative modeling approaches~\cite{yao2023conditional,hu2024accurate,alberts2023learning}, NMR-to-Structure~\cite{alberts2023learning} serves as a representative framework that encodes NMR spectra into token sequences and employs a sequence-to-sequence architecture to generate molecular SMILES~\cite{weininger1988smiles}. While these end-to-end frameworks eliminate the need for manual feature engineering, they suffer from limited generalization and poor interpretability. Moreover, due to the scarcity of high-quality experimental datasets, such models are typically trained on simulated data, leading to a significant domain gap between training and real-world conditions. This undermines their reliability in practical applications and their ability to generalize to out-of-distribution chemical structures.

In contrast, traditional optimization strategies—such as genetic algorithms~\cite{brown2019guacamol,tripp2023genetic}—offer greater transparency by explicitly exploring molecular space through iterative refinement~\cite{jensen2019graph,mirza2024elucidating}. To obtain a suitable initial population, database lookups are commonly employed to rapidly select from known compounds~\cite{burns2019role,yang2021cross,sun2024cross}. 
However, their stochastic and undirected search mechanisms lead to inefficient and unstable navigation of the vast chemical space, often failing to converge to the desired molecules within feasible computational time. These challenges are further exacerbated when addressing the single-molecule objective of structure elucidation.

Despite their differing paradigms, both generative and traditional approaches fail to reliably bridge NMR spectra with novel, chemically valid structures under practical conditions—either due to poor generalization and opacity, or inefficiency arising from random exploration.
There remains a critical need for more effective strategies that fully leverage NMR spectral information and demonstrate robustness and reliability in practical applications.

To address these challenges, we propose NMR-Solver, an automated and interpretable framework for determining small organic molecule structures from $^1$H and $^{13}$C NMR spectra. Built on a physics-guided, fragment-based optimization strategy, NMR-Solver integrates large-scale spectral matching with atomic-level structure–spectrum correlations to guide molecular assembly in a chemically meaningful and transparent manner. Unlike traditional approaches that rely on random and undirected search, our method navigates chemical space in a targeted manner by evolving molecular fragments based on observable spectral features. This enables robust and efficient structure elucidation in real-world scenarios—where novel scaffolds, spectral ambiguities, and incomplete data pose significant challenges—while ensuring full traceability of each structural decision to experimental evidence. As a result, NMR-Solver provides a reliable and trustworthy solution for practical structure determination.

\section{Results} \label{sec:results}

\subsection{Overview of the NMR-Solver framework} \label{subsec:overview}

\begin{figure}[t]
    \centering
    
    \begin{subfigure}[t]{\textwidth}
        \includegraphics[width=\linewidth]{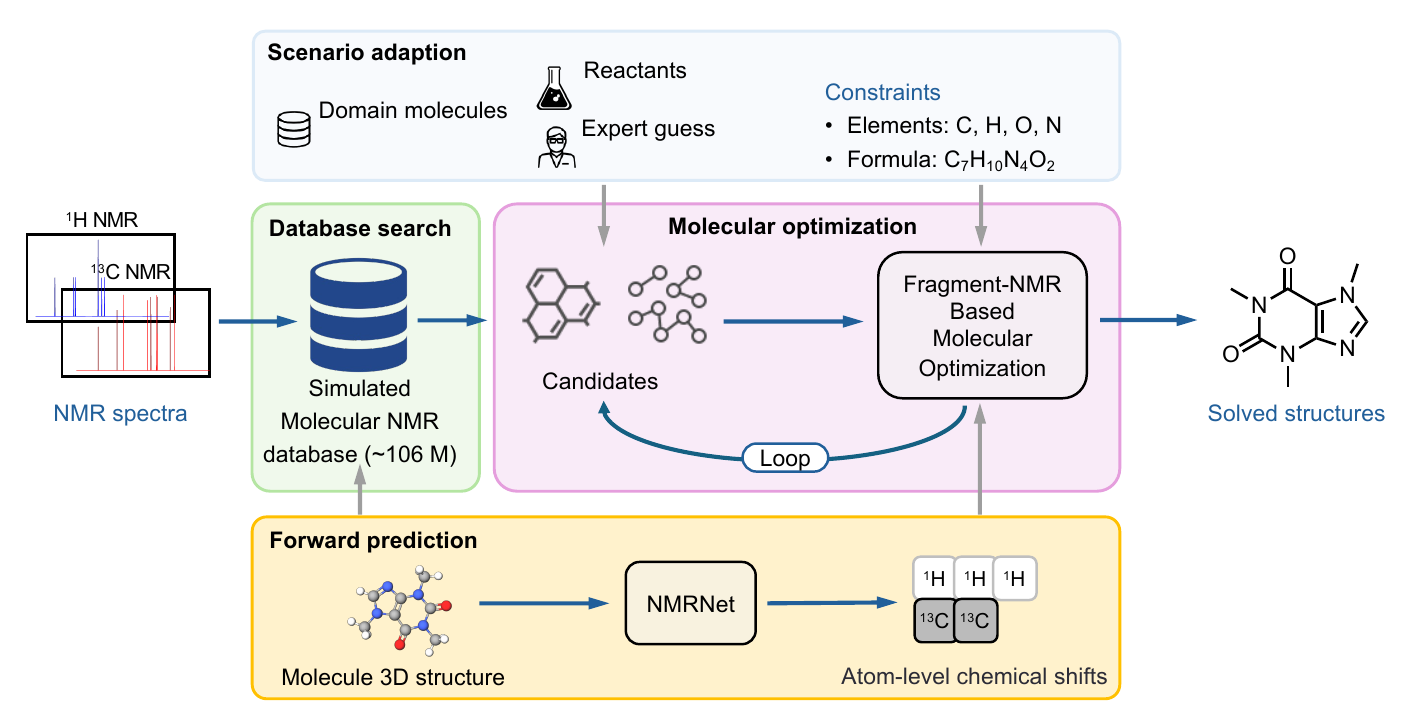}
        \put(-380,190){\textbf{a}} \phantomsubcaption
        \label{fig:framework-a}
    \end{subfigure}
    \hspace{0.05\textwidth}
    \begin{subfigure}[t]{1\textwidth}
        \includegraphics[width=\linewidth]{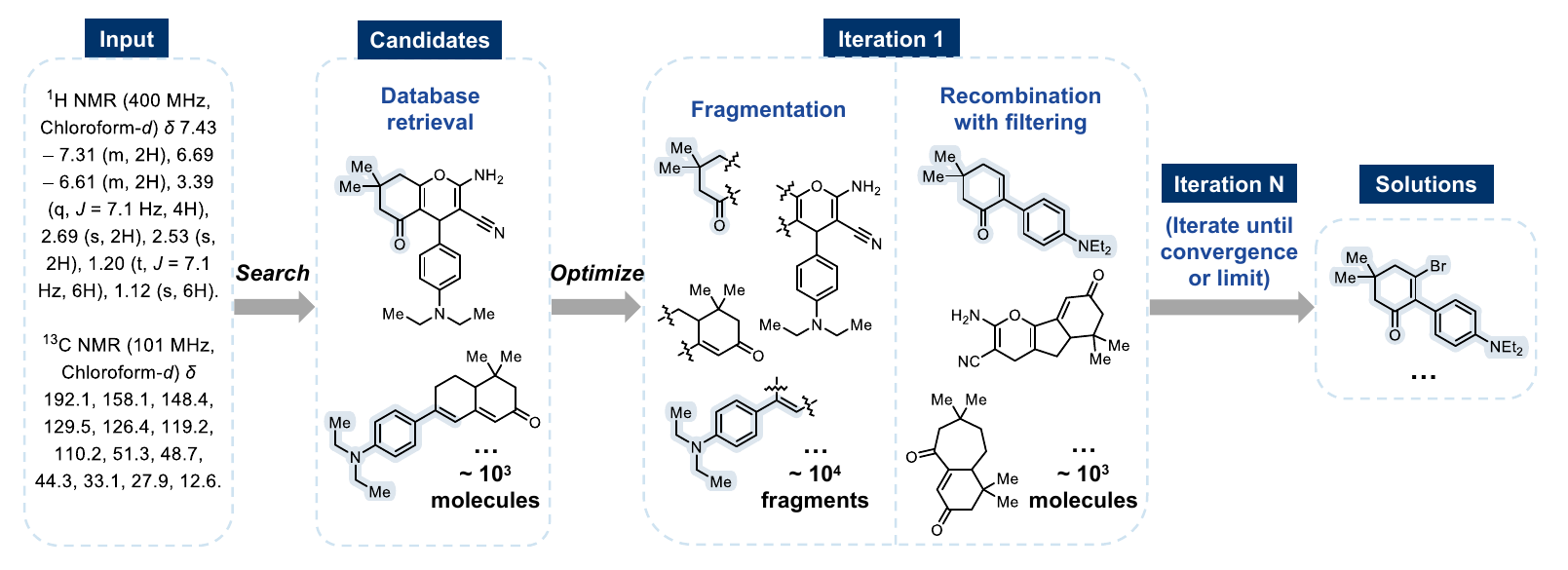}
        \put(-380,130){\textbf{b}} \phantomsubcaption
        \label{fig:framework-b}
    \end{subfigure}

\caption{
    \textbf{a.} \textbf{Architecture of NMR-Solver.} The framework operates in a closed loop and integrates four modules: 
    (i) forward prediction of $^1$H and $^{13}$C chemical shifts using NMRNet; 
    (ii) large-scale database search over 106 million entries; 
    (iii) Fragment-NMR-Based Molecular Optimization (FB-MO) for iterative refinement; and 
    (iv) a scenario adaptation interface allowing user-defined candidates and constraints. 
    \textbf{b.} \textbf{Workflow of NMR-Solver.} Starting from experimental NMR data, candidate molecules are retrieved from a large database. They are fragmented and recombined in an iterative loop, with each iteration filtering new candidates by spectral matching using predicted chemical shifts. The process continues until convergence, producing chemically valid and spectrally consistent solutions.
}
    \label{fig:framework}
\end{figure}

As illustrated in Fig.~\ref{fig:framework-a}, the NMR-Solver framework consists of four core modules: molecular optimization, forward prediction, database retrieval, and scenario adaptation. 
Operating in a closed-loop manner, it addresses the inverse NMR problem by starting from initial candidate structures—obtained via database retrieval or user input—and iteratively refining them through fragment-based molecular optimization. In each iteration, the spectra of evolving candidates are simulated and compared against experimental data, guiding the search toward chemically valid and spectrally consistent solutions (see Fig.~\ref{fig:framework-b} for an illustration of the iterative refinement process).

Central to NMR-Solver is the molecular optimization module, which employs a physics-guided, fragment-based strategy to iteratively refine candidate structures. In each iteration, the module generates new candidates by enumerating chemically valid fragment combinations from the current molecular pool—potentially exceeding $10^9$ possibilities under typical settings. Rather than exploring this space randomly, the search is guided by atomic-level structure–spectrum correlations, enabling efficient filtering of candidates whose simulated spectra best match the experimental data. This focused, evidence-driven exploration ensures both high efficiency and interpretability, with each structural change traceable to specific NMR features.

To support rapid spectral evaluation during optimization, we employ NMRNet~\cite{xu2025toward}, an SE(3)-equivariant Transformer~\cite{vaswani2017attention} architecture that that predicts $^1$H and $^{13}$C chemical shifts with high accuracy (reported MAE: 0.181 ppm for $^1$H, 1.098 ppm for $^{13}$C), comparable to DFT calculations.
Crucially, NMRNet achieves this predictive performance at a computational cost several orders of magnitude lower than DFT, enabling real-time spectral simulation during iterative optimization and large-scale database construction.

For efficient candidate initialization, the framework leverages a large-scale spectral retrieval module that identifies plausible molecular candidates by querying the \textbf{SimNMR-PubChem Database}, which we constructed from approximately 106 million small organic molecules sourced from PubChem~\cite{kim2025pubchem}. Each molecule in this database is annotated with NMRNet-predicted chemical shifts, making it the largest simulated NMR database available to date. This substantially exceeds the scale of existing publicly available datasets, such as NMRShiftDB2~\cite{kuhn2015nmrshiftdb2}, NP-MRD~\cite{wishart2022np}, QM9-NMR~\cite{gupta2021qm9nmr}, and Multimodal Spectroscopic Dataset~\cite{alberts2024unraveling}. By leveraging modern vector database techniques—including approximate nearest neighbor search (ANN)—and applying re-ranking strategies, the database retrieves spectrally similar candidates in sub-second time, providing high-quality starting points for molecular optimization.

Finally, the scenario adaptation module supports the integration of domain-specific knowledge—such as known reactants or proposed scaffolds—as initial inputs for guided structural search. It also allows constraints on molecular formula and elemental composition to be applied during optimization and filtering. Together, these capabilities enable flexible deployment across diverse scenarios, ranging from \textit{ab initio} structure elucidation to reaction-aware product inference.

To promote accessibility, we provide a user-friendly web application~\cite{NMRWeb} for interactive molecular structure elucidation from NMR spectra. 
The platform also provides integrated tools for database search, chemical shift prediction and spectral matching—enabling both automated analysis and human-in-the-loop interpretation.

\subsection{Structure elucidation on simulated spectra}

Due to the scarcity of large-scale, high-quality experimental NMR datasets with carefully curated structure-spectrum pairs, the majority of existing computational methods are evaluated on simulated data. To facilitate a fair comparison with prior work, we evaluated NMR-Solver on the publicly available benchmark dataset introduced by Alberts et al.~\cite{alberts2023learning}. This dataset has been used to benchmark several state-of-the-art approaches, including NMR-to-Structure~\cite{alberts2023learning} and GraphGA with Multimodal Embeddings~\cite{mirza2024elucidating}, whose performance results are reported in the original study. We compare our method directly against these approaches, enabling a meaningful assessment of NMR-Solver’s capabilities within the context of current state-of-the-art techniques.

The dataset comprises 345,000 SMILES–spectrum pairs, with $^1$H and $^{13}$C NMR spectra simulated using MestReNova~\cite{Mestre}, where each $^1$H spectrum includes chemical shifts, peak integrations, $J$-coupling constants, and multiplicity patterns, and each $^{13}$C spectrum contains chemical shifts. To assess performance on this simulated benchmark, we constructed a test set by randomly sampling 1,000 molecules from the dataset. Fluorine-containing compounds were excluded from evaluation due to inconsistent handling of C–F coupling in the simulation pipeline—specifically, split resonances were modeled as multiple discrete peaks, leading to spectral misalignment that could introduce bias in model assessment.

We evaluated the method using the original metrics, requiring an exact molecular match with stereochemistry considered, and the results are summarized in Table~\ref{tab:result_simulated}. Although our approach does not use the same simulation pipeline as (MestReNova) GraphGA with Multimodal Embeddings, nor is NMR-Solver trained on the benchmark dataset like NMR-to-Structure, it still achieves performance comparable to state-of-the-art methods on both $^{13}$C-only and combined $^1$H \& $^{13}$C tasks. The performance gap in the $^1$H-only setting arises because previous methods rely on $^1$H multiplicity patterns and $J$-coupling values from simulated spectra—idealized signals that are rarely fully or accurately observable in real experimental conditions due to solvent effects, peak overlap, impurity peaks, and instrument resolution. Consequently, earlier models may overfit these idealized features in the simulated $^1$H NMR spectra, whereas our method focuses on signals that are reliably measurable in practice.

This indicates that NMR-Solver effectively leverages experimentally relevant spectral features for accurate molecular structure prediction, rather than relying on fragile information such as idealized multiplicity patterns or artificially noise-free simulations. In the following section, we further evaluate the framework on experimentally measured NMR spectra curated from the literature, highlighting its generalization capability and practical robustness in real-world structure elucidation tasks.

\begin{table}[h]
    \caption{Comparison with current methods on the simulated dataset. The results for \textbf{GraphGA with Multimodal Embeddings} and \textbf{NMR-to-Structure} are directly reported from the respective papers. Previous methods benefit from idealized simulated spectra, which are rarely fully reproduced in real experiments. Results in \textbf{bold} indicate the best performance for each setting, while results with \underline{underlines} indicate the best performance across all settings.}
    \label{tab:result_simulated}
    
    \begin{tabular*}{\textwidth}{@{\extracolsep{\fill}} l l c c c }
    \toprule
        Input Spectra & Conditions & Top-1 (\%) & Top-5 (\%) & Top-10 (\%) \\
    \midrule
        \multicolumn{5}{l}{\textbf{GraphGA with Multimodal Embeddings~\cite{mirza2024elucidating}}} \\
    \midrule
        $^1$H NMR + $^{13}$C NMR + IR & Formula & \underline{76.02} & \underline{87.81} & 88.91 \\
    \midrule
        \multicolumn{5}{l}{\textbf{NMR-to-Structure~\cite{alberts2023learning}}} \\
    \midrule
        $^1$H NMR                & Formula & \textbf{55.32} & \textbf{73.59} & \textbf{76.74} \\
        $^{13}$C NMR             & Formula & 53.91 & 73.45 & 77.72 \\
        $^1$H NMR + $^{13}$C NMR & Formula & \textbf{66.99} & 84.09 & 86.59 \\
    \midrule
        \multicolumn{5}{l}{\textbf{NMR-Solver (Ours)}} \\
    \midrule
        $^1$H NMR & Formula & 23.10 & 36.20 & 39.30 \\
        $^{13}$C NMR & Formula & \textbf{62.20} & \textbf{77.60} & \textbf{79.10} \\
        $^1$H NMR + $^{13}$C NMR & Formula & 66.90 & \textbf{87.20} & \underline{\textbf{89.90}} \\
    \bottomrule
    \end{tabular*}
\end{table}

\subsection{Generalization to experimental spectra}

To enable realistic and rigorous benchmarking of structure elucidation methods, we manually curated a dataset of approximately 450 reactant–product pairs with product NMR spectra from original research articles published in the \textit{Journal of the American Chemical Society} (JACS) in 2024. 

The data were extracted from 15 weekly issues, covering a broad range of organic chemistry disciplines, total synthesis, catalysis, and substrate preparations in polymer chemistry and biochemistry. For each article, 3–5 representative reactions were selected to maximize structural and functional diversity. Each entry includes the key reactants, the reported product structure, and its corresponding experimental $^1$H and $^{13}$C NMR spectra. Minor manual curation was performed to remove peaks originating from solvents or impurities, ensuring high data quality. This benchmark reflects real-world conditions in synthetic chemistry and provides a robust foundation for evaluating computational structure elucidation approaches.

A major limitation of existing approaches, such as NMR-to-Structure, lies in their reliance on idealized simulated spectra, which often fail to capture the complexities of real-world measurements. In these datasets, $^1$H NMR spectra are typically generated with fully resolved multiplet patterns and exact $J$-coupling values, assuming ideal conditions that are rarely achieved experimentally. In practice, such information is frequently obscured by linewidth broadening, signal overlap, impurities, and instrument resolution—for example, in our literature-extracted dataset, approximately 41\% of $^1$H NMR peaks are annotated simply as "m" (multiplet), without interpretable coupling constants.

\begin{figure}[htbp]
    \begin{subfigure}[t]{0.45\textwidth}  %
        \includegraphics[width=\linewidth]{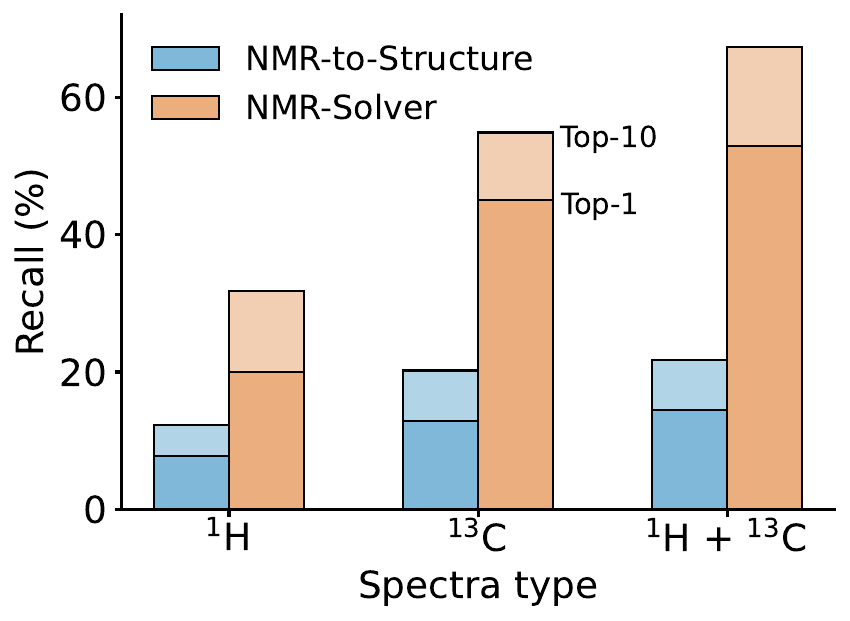}
        \put(-175,115){\textbf{a}} \phantomsubcaption
        \label{fig:result_exp-a}
    \end{subfigure}
    \begin{subfigure}[t]{0.45\textwidth}
        \includegraphics[width=\linewidth]{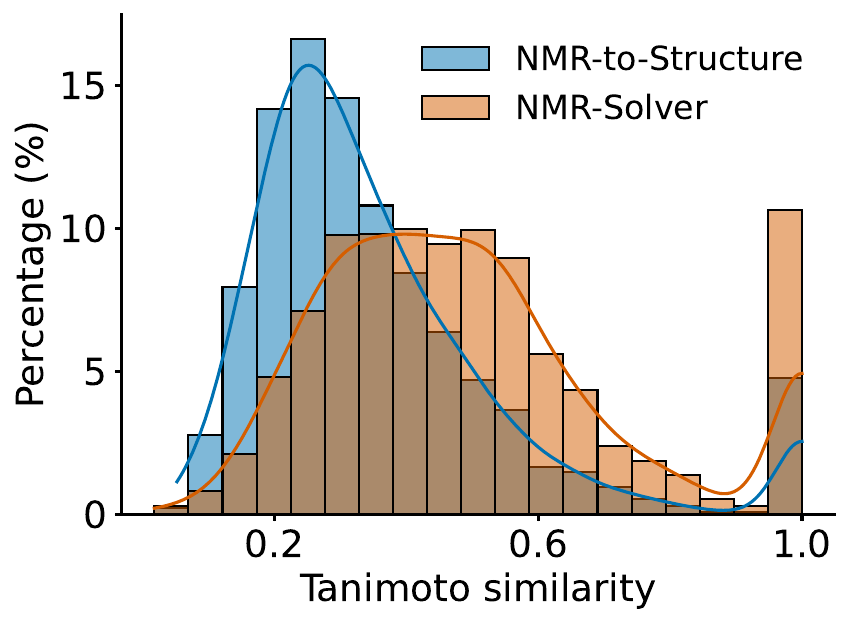}
        \put(-175,115){\textbf{b}} \phantomsubcaption
        \label{fig:result_exp-b}
    \end{subfigure}
    \begin{subfigure}[t]{0.49\textwidth}
        \includegraphics[width=\linewidth]{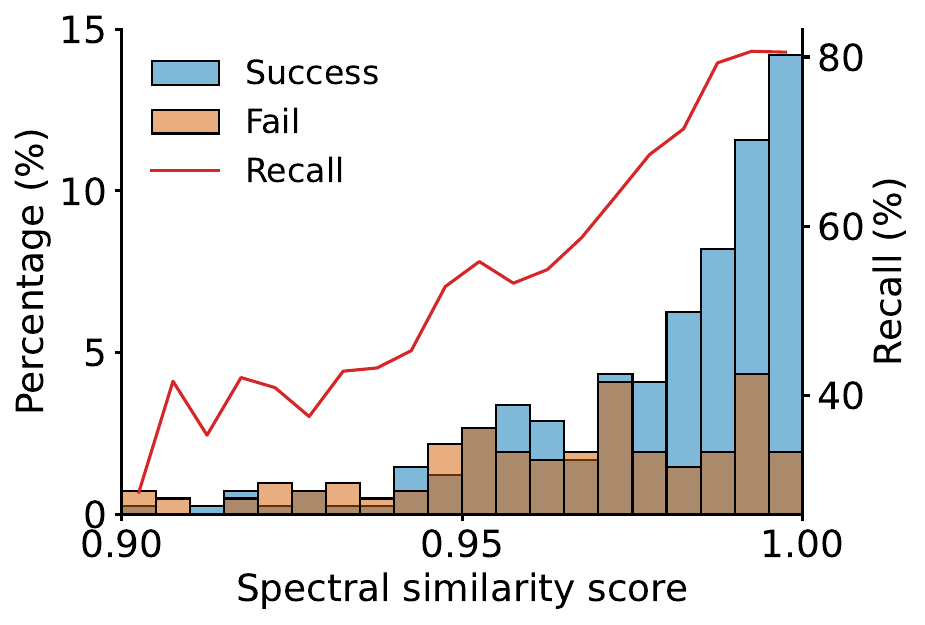}
        \put(-190,115){\textbf{c}} \phantomsubcaption
        \label{fig:result_exp-c}
    \end{subfigure}
    \hspace{0.05\textwidth}
    \begin{subfigure}[t]{0.45\textwidth}
        \includegraphics[width=\linewidth]{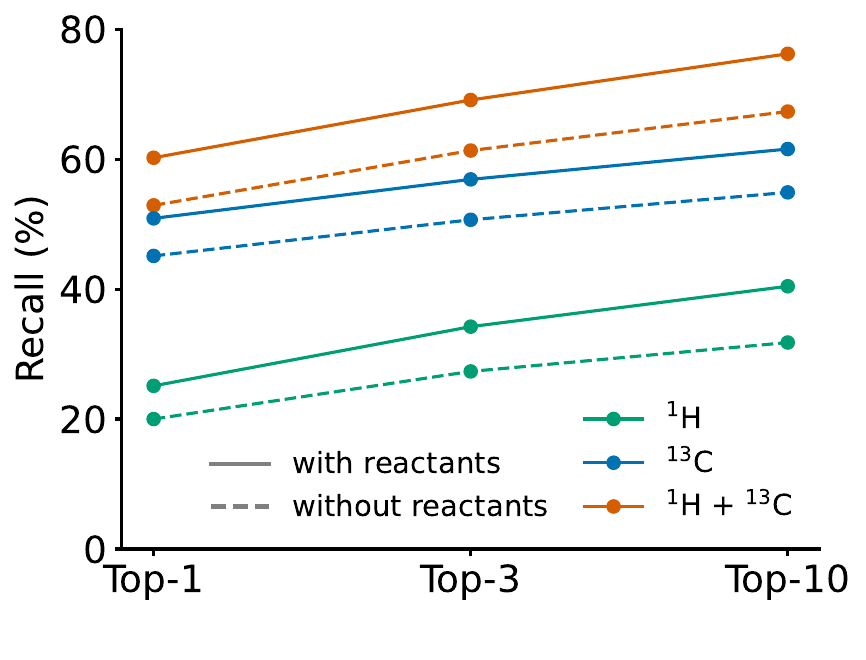}
        \put(-175,115){\textbf{d}} \phantomsubcaption
        \label{fig:result_exp-d}
    \end{subfigure}
    \begin{subfigure}[t]{0.45\textwidth}
        \includegraphics[width=\linewidth]{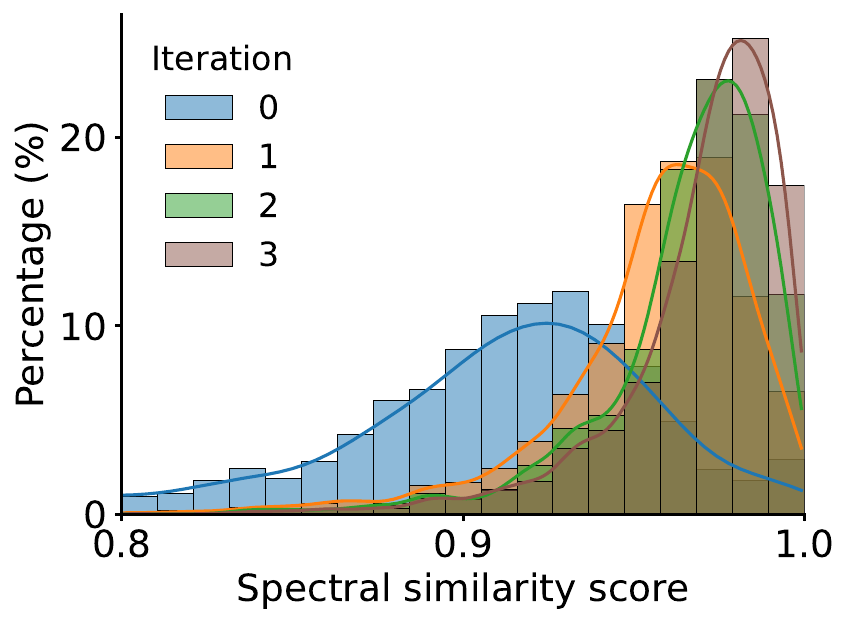}
        \put(-175,115){\textbf{e}} \phantomsubcaption
        \label{fig:result_exp-e}
    \end{subfigure}
    \hspace{0.075\textwidth}
    \begin{subfigure}[t]{0.45\textwidth}
        \includegraphics[width=\linewidth]{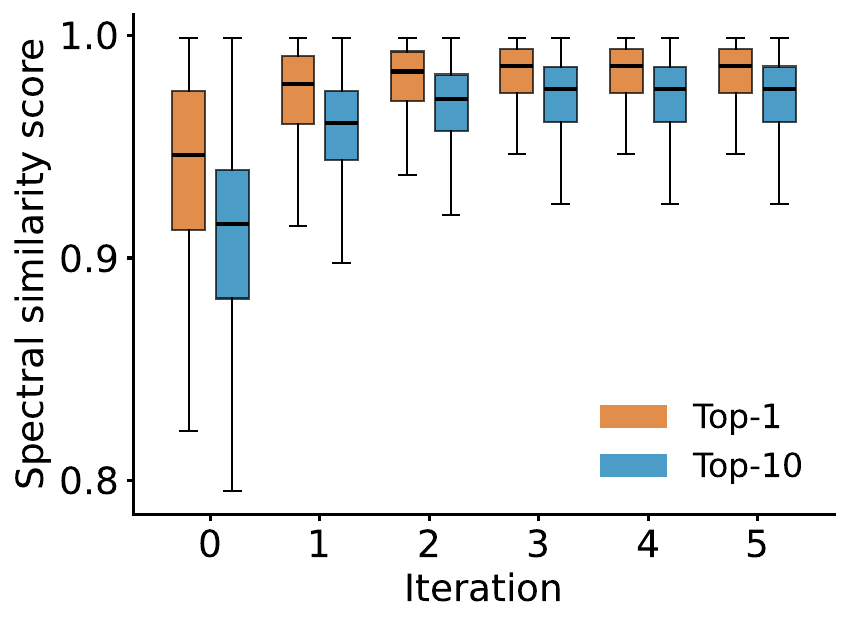}
        \put(-175,115){\textbf{f}} \phantomsubcaption
        \label{fig:result_exp-f}
    \end{subfigure}
    
\caption{\textbf{Performance of structure elucidation on literature data.}
    \textbf{a.} Comparison of top-1 and top-10 recall for different methods under the formula condition.
    \textbf{b.} Comparison of tanimoto similarity distributions for top-10 predictions, excluding invalid predictions and SMILES.
    \textbf{c.} Relationship between average top-10 spectrum similarity score and success rate.
    \textbf{b, c.} Results are conducted by NMR-Solver using $^1$H and $^{13}$C NMR under the formula condition.
    \textbf{d.} Comparison of recall rates for NMR-Solver with and without reactants.
    \textbf{e.} Dynamics of spectral similarity scores across iterations.
    \textbf{f. } Top-1 and top-10 score distributions are shown as box plots (median, IQR, and 10th–90th percentiles).
}
    \label{fig:result_exp}
\end{figure}

We evaluated NMR-Solver and NMR-to-Structure using both recall and Tanimoto similarity~\cite{bajusz2015tanimoto} computed on Morgan fingerprints~\cite{morgan1965generation}, which measure exact structure recovery and substructural similarity between predicted and ground-truth molecules, respectively. As shown in Fig.~\ref{fig:result_exp-a}, The earlier method NMR-to-Structure exhibits limited performance, achieving only 14.44\% (top-1) and 21.78\% (top-10) accuracy when using $^1$H and $^{13}$C NMR spectra.
In contrast, NMR-Solver substantially outperforms these baselines, achieving 52.89\% (top-1) and 67.33\% (top-10) recall under the same evaluation conditions. Moreover, as shown in Fig.~\ref{fig:result_exp-b}, even for predictions that do not exactly match the ground truth, NMR-Solver yields structures with higher Tanimoto similarity, reflecting improved substructural consistency and chemically meaningful outputs.

These results highlight a substantial performance gap when models trained on simulated data are applied to real-world, experimental measured NMR spectra. In contrast, our method does not rely on large-scale simulated data for training, but instead leverages simulation only during the scoring stage, leading to greater robustness in practical scenarios.

Furthermore, NMR-Solver ranks candidate structures by spectral similarity between simulated and experimental spectra. As illustrated in Fig.~\ref{fig:result_exp-c}, prediction accuracy improves monotonically with increasing spectral similarity, suggesting that the spectral match score can serve as a reliable confidence indicator for the predictions. This self-consistent scoring mechanism enhances the reliability and practical applicability of NMR-Solver in real-world structure elucidation tasks.

To facilitate comparison with previous work, stereoisomers are ignored in this evaluation.
A complete breakdown of experimental results is provided in Supplementary Table~\ref{tab:result_exp_wo_stereo} and~\ref{tab:result_exp_with_stereo}.

\subsection{Product structure prediction guided by reaction context}

In practical structural analysis, chemists often interpret NMR spectra in the context of known reactants or anticipated chemical transformations. 
To evaluate whether NMR-Solver can similarly benefit from such prior knowledge, we incorporated reactant information from the literature-extracted dataset described above.

By introducing reactant structures into the candidate pool via the scenario adaptation module, NMR-Solver leverages substructures that are likely retained in the product and reflected in the NMR spectra. This enables the method to effectively bias molecular optimization toward candidates containing these spectrally supported, chemically plausible fragments.

As illustrated in Fig.~\ref{fig:result_exp-d}, this strategy increases the top-1 recall from 52.89\% to 60.22\% and the top-10 recall from 67.33\% to 76.22\% under $^1$H and $^{13}$C NMR conditions with molecular formula provided, demonstrating that leveraging structural continuity across reactions enhances inference accuracy.

These results highlight how NMR-Solver combines automated analysis with human-like reasoning: rather than operating in isolation, it amplifies expert knowledge and integrates reaction information in a flexible and robust manner, improving both the reliability and general applicability of molecular structure elucidation in synthetic workflows. 
While our evaluation focuses on incorporating reactant structures, the framework is generalizable and can similarly leverage expert hypotheses or domain-specific molecular priors to guide optimization when available.

\subsection{Evolution of spectral similarity during molecular optimization}

During molecule optimization process, the agreement between predicted and experimental NMR spectra evolves as candidate molecules are iteratively refined. 
As shown in Fig.~\ref{fig:result_exp-e} and Fig.~\ref{fig:result_exp-f}, spectral similarity scores for top candidates increase substantially within the first few optimization steps, indicating rapid convergence toward structures with high spectral agreement.

Specifically, the median spectral similarity of the top-10 candidates surpasses 0.96 within the first two iterations—a threshold empirically associated with ~50\% top-10 prediction accuracy—and gradually approaches a plateau of ~0.97. The diminishing gains in later stages reflect that the search quickly homes in on chemically meaningful structures.

This rapid progression underscores the efficiency of NMR-Solver in navigating the vast chemical space: by prioritizing molecular structures that better explain the observed spectra, the method identifies high-quality candidates early without exhaustive exploration. In contrast, conventional genetic algorithm–based approaches rely on largely random crossover and mutation operations, lacking directed guidance. These methods explore chemical space indiscriminately, often requiring many more iterations to converge and producing candidates of highly variable quality.

Overall, the progressive refinement of spectral match illustrates a coherent, goal-directed search process, where candidate quality systematically improves at each step, highlighting both the effectiveness and practicality of the optimization strategy for real-world applications.

\subsection{Real-world experimental validation}

\begin{figure}[H]
    \centering
    \includegraphics[width=1\linewidth]{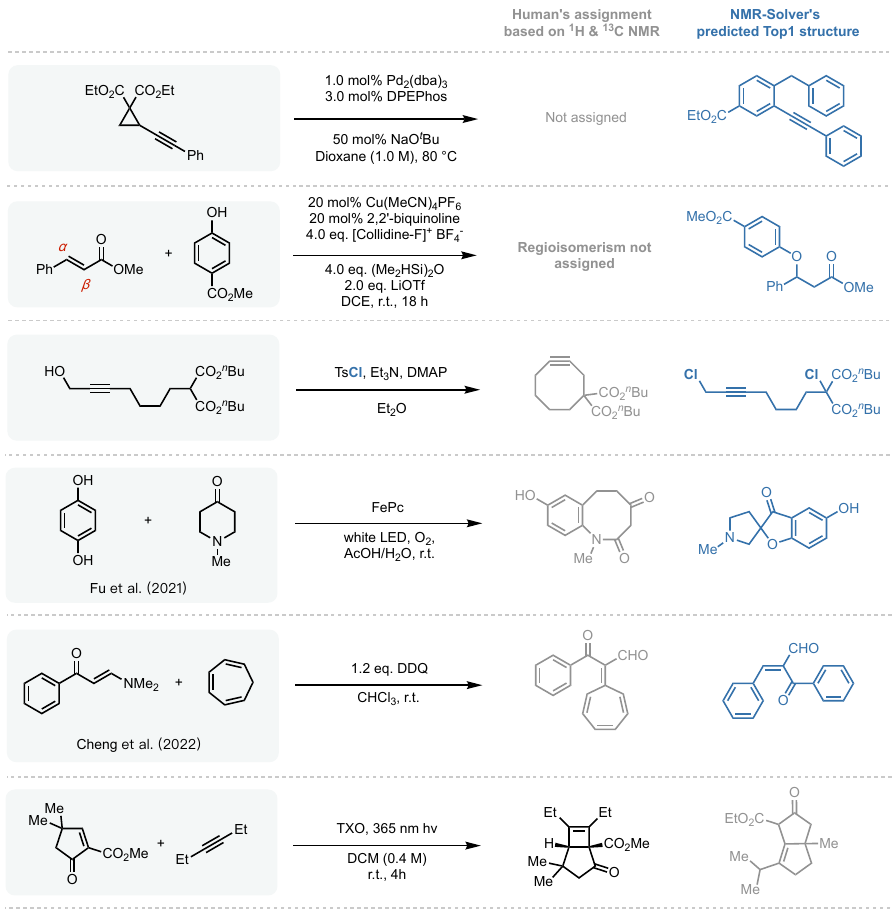}
    \put(-380,345){\textbf{a}}
    \put(-380,289){\textbf{b}}
    \put(-380,233){\textbf{c}}
    \put(-380,177){\textbf{d}}
    \put(-380,111){\textbf{e}}
    \put(-380,45){\textbf{f}}
\caption{
\textbf{Real-world experimental cases demonstrating the utility of NMR-Solver.} 
\textbf{a.} Challenging laboratory case where manual NMR analysis failed but NMR-Solver correctly predicted the structure.  
\textbf{b.} Regioisomer discrimination in cinnamate ester hydrofunctionalization, correctly resolved by NMR-Solver.  
\textbf{c.} Unanticipated side product lacking aromatic protons, correctly proposed as a dichlorinated structure by NMR-Solver.  
\textbf{d–e.} Correction of two misassigned structures in the literature using NMR-Solver. 
\textbf{f.} Failing to predict the correct structure due to inaccuracies in the forward prediction model.
}
    \label{fig:real-world}
\end{figure}
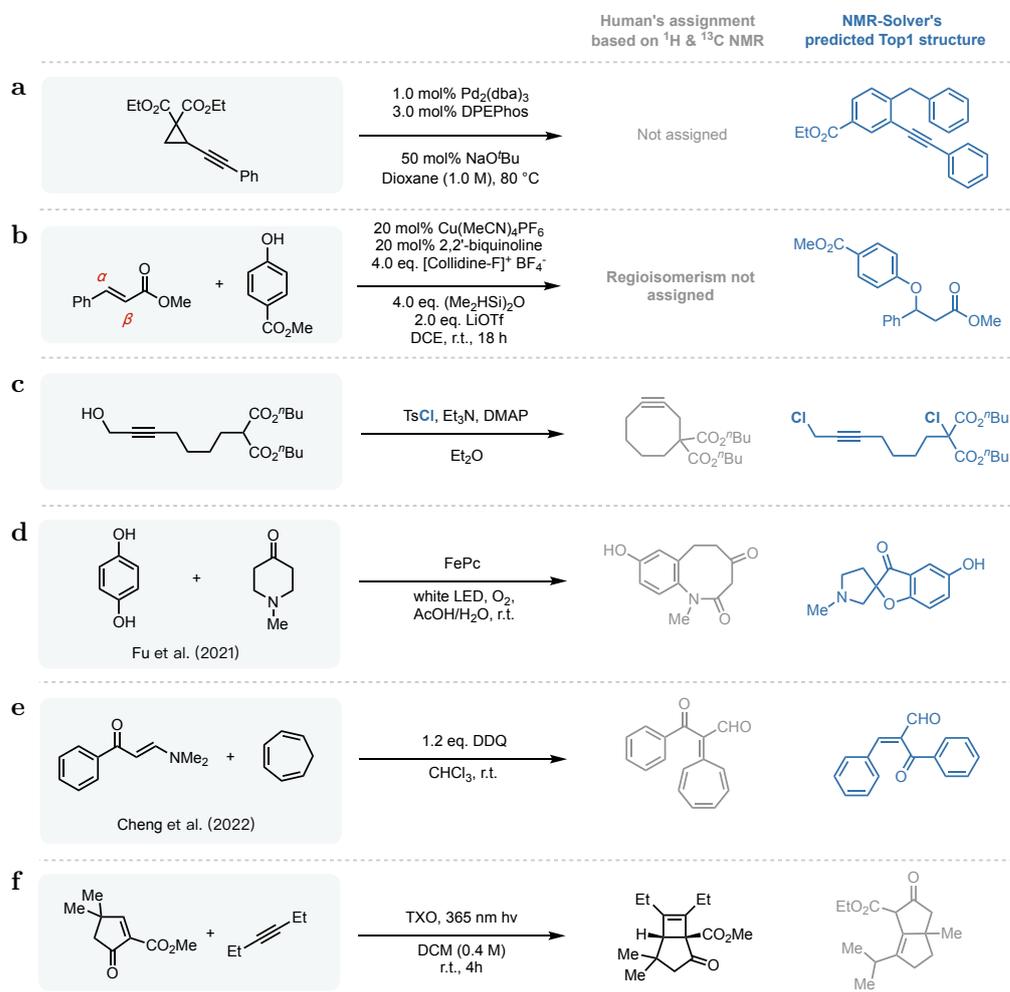

To further assess the practical utility of NMR-Solver in real-world synthetic chemistry research, we applied it to challenging cases from laboratory experiments—situations where traditional manual analysis is hard to provide conclusive structural assignments. Representative cases from real-world experiments are presented in Fig.~\ref{fig:real-world}.

In some instances, reaction pathways are highly complex or unanticipated, making it difficult for chemists to hypothesize plausible product structures based on mechanistic reasoning alone; without such hypotheses, manual NMR interpretation becomes extremely challenging. 
An illustrative example (Fig.~\ref{fig:real-world}a) from our laboratory involved a compound for which initial $^1$H and $^{13}$C NMR spectra did not allow reliable structure assignment, and the corresponding study was temporarily halted. 
Applying NMR‑Solver to the same NMR data enabled identification of a candidate structure that consistently matched the observed chemical shifts and coupling patterns. Subsequent validation via two-dimensional NMR experiments (HMBC and HSQC) and high-resolution mass spectrometry (HRMS) confirmed the correctness of the assignment.

Isomeric products further complicate structural elucidation, as many isomers produce very similar $^1$H and $^{13}$C spectra, making them difficult to distinguish without additional references.
For example, In the case of hydrofunctionalization of cinnamate esters (Fig.~\ref{fig:real-world}b), two regioisomers are possible. Manual interpretation of the one-dimensional spectra without references is difficult to distinguish the products. NMR‑Solver, however, accurately discriminated between the isomers based solely on the one‐dimensional spectra, correctly identifying the site of hydrogenation and thereby corroborating the proposed reaction mechanism.

Unexpected side products pose additional challenges. For instance, in an attempt to convert a hydroxyl group into a sulfonate ester (Fig.~\ref{fig:real-world}c), chromatographic isolation yielded a byproduct lacking any aromatic protons. Initial hypotheses considered included chlorination of the hydroxyl group, which would introduce one additional proton compared with the observed spectrum, or the formation of an eight-membered cyclic byproduct; however, HRMS analysis showed no mass corresponding to the eight-membered cyclic structure. When NMR‑Solver was applied, it confidently proposed a dichlorinated product whose predicted spectrum matched the experimental data, and subsequent HRMS confirmed the molecular formula of the dichloride, validating this assignment.

Beyond facilitating elucidation of previously unassignable compounds, NMR-Solver can detect and correct misassigned structures in the literature. Two cases (Fig.~\ref{fig:real-world}d, \ref{fig:real-world}e) originally reported by Fu \textit{et al.}~\cite{fu2021photocatalyzed} and Cheng \textit{et al.}~\cite{cheng2022ddq}, and later corrected by Novitskiy \textit{et al.}~\cite{novitskiy2022peculiar}, were re-analyzed with NMR-Solver. The corrected structures were accurately predicted and received substantially higher spectral similarity scores than the originally proposed ones, demonstrating that NMR-Solver can independently identify inconsistencies that might elude manual interpretation, even in peer-reviewed publications.

A key limitation of NMR‑Solver lies in its reliance on the accuracy of the underlying forward prediction model.
In one example involving a strained system (Fig.~\ref{fig:real-world}f), the prediction model (NMRNet) exhibited relatively large deviation, leading NMR‑Solver to favor a structure with a predicted spectrum closer to the experimental data rather than the real product. 
Future advances and anticipated improvements in NMR prediction algorithms will be incorporated into NMR-Solver to continuously advance its accuracy in structural determination.

\section{Discussion}

NMR-Solver enables accurate, robust, and interpretable structure elucidation directly from experimental $^1$H and $^{13}$C NMR spectra. Unlike many existing methods that treat structure elucidation as a black-box translation task, NMR-Solver constructs molecules through chemically meaningful steps guided by spectral evidence, making interpretability intrinsic to the generation process. At the same time, it achieves state-of-the-art performance and strong robustness against common spectral ambiguities, supporting reliable application under realistic experimental conditions.

Looking ahead, two directions appear particularly important for advancing the field. First, systematic curation of large-scale experimental NMR datasets—from literature mining or high-throughput measurements—will be critical for training and benchmarking methods under practical conditions. Second, continued progress in forward spectral simulation, especially in accurate chemical shift prediction, will further enhance the reliability and generalizability of inference. With its modular design, NMR-Solver is well positioned to leverage improvements in both dataset availability and spectral simulation.

By integrating physics-informed structure–spectrum relationships into the optimization process, NMR-Solver establishes a guided and robust paradigm for molecular optimization, moving beyond the inefficiencies and instability of traditional stochastic approaches. This framework also holds promise for extension to other spectroscopic modalities and broader molecular design applications.

Finally, as an interpretable and flexible analysis platform, NMR-Solver accommodates both fully automated and human-in-the-loop spectrum interpretation. In the context of emerging automated laboratories, it can serve as a key component for validating reaction outcomes and inferring products, thereby accelerating the discovery of new reactions and synthetic pathways. From a broader perspective, such capabilities underscore its potential role in driving a paradigm shift toward automation-assisted scientific research.

\section{Methods}

\subsection{Data preparation}

NMR spectra are processed into a list of peaks annotated with relevant features. For $^1$H NMR, each peak includes chemical shift, integration (proton count), multiplicity, and $J$-coupling constants; for $^{13}$C NMR, only the chemical shift is retained.

For literature data, features are extracted from textual reports using regular expressions from textual NMR reports. For experimental spectra, peaks are annotated using software such as MestReNova~\cite{Mestre}, typically via manual assignment or assisted automation.

In experimental $^1$H and $^{13}$C NMR spectra, the chemical shifts of individual peaks are extracted and represented as an unordered multiset, serving as a unified spectral representation. To encode integration information in $^1$H NMR, each chemical shift is repeated in the multiset according to its associated proton count. Multiplicity are additionally attached to each shift. When a chemical shift is reported as a range, the midpoint of the range is taken as the representative value.

\subsection{Spectral simulation} \label{subsec:method-simulation}

Chemical shift predictions are generated using NMRNet, a deep learning model that takes atomic types and three-dimensional coordinates as input and predicts chemical shifts of all $^1$H and $^{13}$C nuclei. Prior to prediction, molecular conformations are generated using the \texttt{EmbedMolecule} and \texttt{MMFFOptimizeMolecule} functions from the RDKit toolkit~\cite{landrum2016rdkit}, which perform conformational sampling and geometry optimization under the Merck Molecular Force Field (MMFF)~\cite{halgren1996merck}.

The final simulated spectra are also encoded as unordered multisets: $ \mathcal{H} = \{h_1, \dots, h_n\} $ for $^1$H NMR, where each element $h_i$ corresponds to the predicted chemical shift of an individual hydrogen nucleus, and $ \mathcal{C} = \{c_1, \dots, c_m\} $ for $^{13}$C NMR, where each element $c_j$ corresponds to a unique carbon signal, with chemical shifts from symmetry-equivalent carbon nuclei averaged into a single resonance. For $^{13}$C NMR, chemical shifts of symmetry-equivalent carbons are averaged into a single resonance. This representation naturally captures signal intensity through the multiplicity of identical shift values arising from chemically equivalent nuclei, and mimics the peak coalescence observed in experimental spectra.

\subsection{Similarity metric of NMR spectra} \label{subsec:method-similarity}

To enable efficient and robust scoring between two NMR spectra, we introduce two complementary similarity metrics: \textbf{vector similarity} designed for scalable retrieval and high-throughput screening, and \textbf{set similarity} tailored to improve accuracy and robustness in the presence of spectral noise and peak mismatches.

\subsubsection*{Vector similarity}

Given an NMR spectrum $ \mathcal{X} = \{x_1, \dots, x_n\} $, we smooth it using Gaussian convolution:
\begin{equation}
    g(t) = \sum_{i=1}^{n} \exp\left( - \frac{(t - x_i)^2}{2\sigma^2} \right),
\end{equation}
where $ t $ spans the typical chemical shift range for $^1$H or $^{13}$C NMR. The resulting signal is discretized into a vector representation:
\begin{equation}
    \mathbf{v}_{\text{X}} = [g(t_1), g(t_2), \dots, g(t_{128})],
\end{equation}
with $ \{t_i\} $ denoting 128 uniformly spaced sampling points across the spectral window. The specific parameter settings are provided in Supplementary Table~\ref{tab:scoring_parameters}.

The \textbf{encoding vector} of the full NMR spectrum is formed by concatenating $^1$H and $^{13}$C NMR representations:
\begin{equation} \label{eq:encode}
    \mathbf{v} = [\mathbf{v}_{\text{H}}; \mathbf{v}_{\text{C}}] \in \mathbb{R}^{256},
\end{equation}
yielding a fixed-dimensional representation suitable for fast similarity scoring and scalable vector indexing. Spectral similarity is then measured using the Euclidean distance:
\begin{equation}
    d_{\mathrm{L}^2}(\mathbf{v}, \mathbf{v}') = \|\mathbf{v} - \mathbf{v}'\|_2,
\end{equation}
providing a computationally efficient metric for high-throughput screening.

\subsubsection*{Set similarity}

We define set similarity as the optimal alignment between two collections of chemical shifts, formulated as a bipartite matching problem. Let  
\begin{equation}
    \mathcal{X} = \{x_1, \dots, x_m\}, \quad 
    \mathcal{Y} = \{y_1, \dots, y_n\}
\end{equation}
denote two spectra. The similarity score is defined as
\begin{equation}  \label{eq:bipartite}
    S(\mathcal{X}, \mathcal{Y}) = \frac{1}{\sqrt{mn}} \max_{P \in \mathcal{P}} \sum_{(i,j) \in P} f(x_i, y_j),
\end{equation}
where $ \mathcal{P} $ denotes the set of all one-to-one matchings (allowing unmatched elements), and $ f(x, y) $ is a kernel function measuring pairwise compatibility:
\begin{equation}
    f(x, y) = \exp\!\left( - \frac{(x - y)^2}{2\sigma^2} \right).
\end{equation}

The optimization problem in equation~\eqref{eq:bipartite} can be fast solved using the Kuhn–Munkres algorithm~\cite{kuhn1955hungarian, Munkres1957hungarian}, as implemented in \texttt{linear\_sum\_assignment} from \texttt{scipy.optimize}~\cite{crouse2016implementing}. 
In addition, multiplicity patterns can be incorporated by weighting the kernel function, thereby introducing additional spectral information (see Supplementary Note~\ref{sec:multiplet-type} for implementation details). 

Compared to vector similarity metric, set similarity offers higher tolerant to missing or noisy peaks, and enables robust identification of molecules with partial spectral overlap. Additional experimental evidence demonstrating these advantages is provided in Supplementary Note~\ref{sec:similarity}.

\subsection{Database construction and retrieval} \label{sebsec:method-database}

We constructed the SimNMR-PubChem Database, a large-scale NMR molecular repository derived from the PubChem dataset~\cite{kim2025pubchem}, which initially contained approximately 119 million compounds. To ensure chemical relevance, ionized species, molecules lacking both carbon and hydrogen, free radicals, and isotopically labeled compounds were excluded. After deduplication using InChIKeys, the final dataset comprises 106 million unique molecules.

For each molecule, we generated simulated $^1$H and $^{13}$C NMR spectra as described in Section~\ref{subsec:method-simulation}, and subsequently converted them into 256-dimensional encoding vectors as defined in Eq.~\ref{eq:encode}.

To enable scalable similarity search, we implemented a vector database using FAISS~\cite{johnson2019billion}, employing an HNSW index~\cite{malkov2018efficient} with cosine similarity for efficient approximate nearest neighbor (ANN) retrieval. 

Retrieval is performed in two stages: an initial ANN scan identifies candidate molecules based on vector similarity, followed by re-ranking using the more accurate but computationally slower set similarity. This hybrid approach combines computational efficiency with high-precision spectral matching, enabling sub-second retrieval.

\subsection{Fragment-NMR-Based Molecular Optimization} \label{subsec:FB-MO}

\begin{figure}[t]
    \centering
    \includegraphics[width=\linewidth]{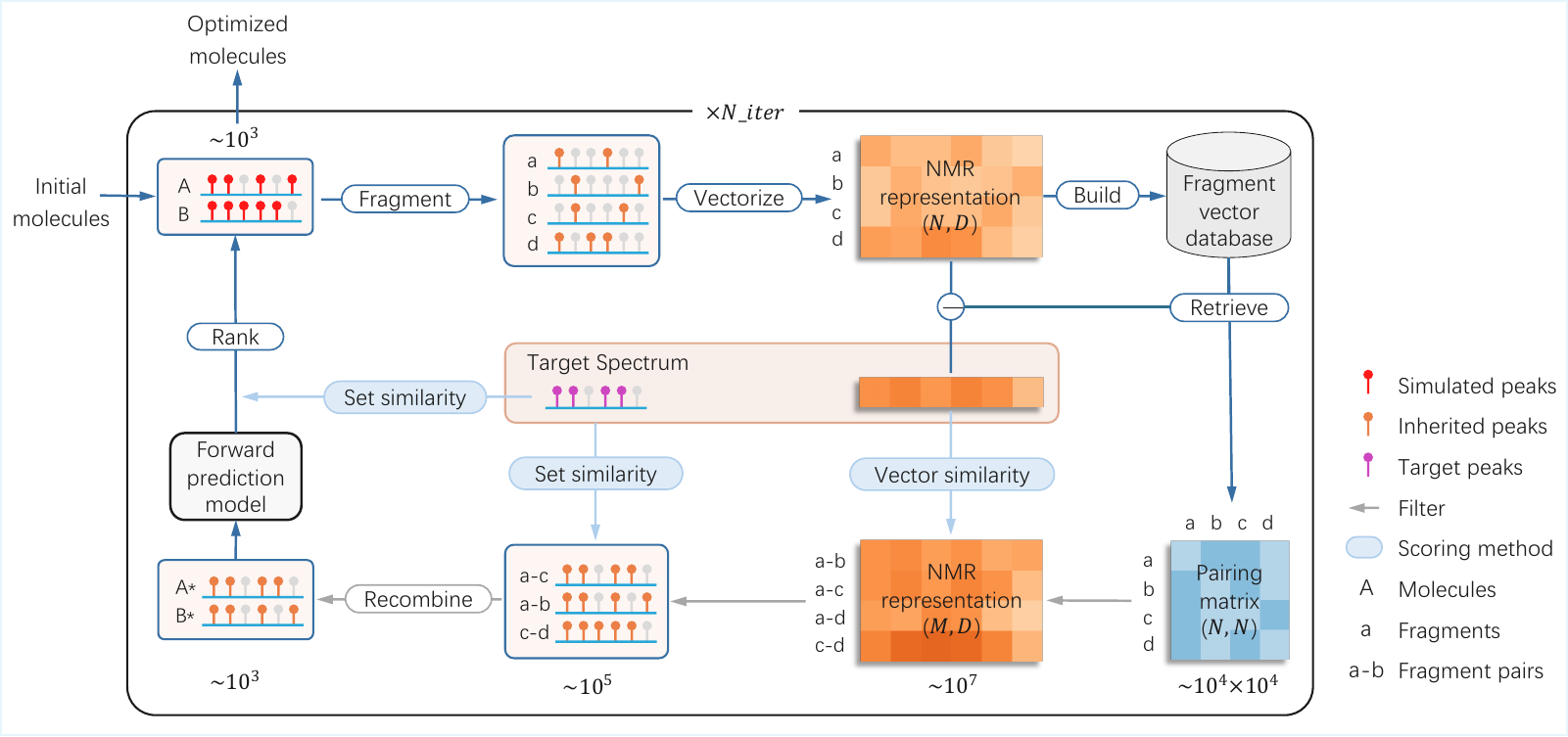}
    \caption{\textbf{Workflow of Fragment-NMR-Based Molecular Optimization (FB-MO).} Array shapes are indicated in parentheses: $N$, number of fragments; $M$, number of selected fragment pairs; $D$, dimensionality of the vector representation.}
    \label{fig:FB-MO}
\end{figure}

We introduce Fragment-NMR-Based Molecular Optimization (FB-MO), a directed evolutionary strategy that leverages predicted NMR spectra to guide the search toward molecules matching a target spectrum. The overall workflow is summarized in Fig.~\ref{fig:FB-MO}.

The approach maintains a dynamically updated molecule pool, initially populated with compounds retrieved from the database (Section~\ref{sebsec:method-database}). In each iteration, these molecules are fragmented into structural building blocks and recombined through crossover to generate novel candidates.

To efficiently predict the NMR spectra of newly generated molecules after crossover, FB-MO estimates the chemical shifts of atoms in the new molecule by using the corresponding shifts from the parent molecules. This strategy can be seen as a fragment-based extension of the HOSE (Hierarchically Ordered Spherical Environment) code approach~\cite{bremser1978hose}, which predicts chemical shifts based on local chemical topology. Its validity is supported by the well-established principle that NMR chemical shifts are primarily governed by local chemical structure, with long-range effects decaying rapidly and typically being negligible~\cite{keeler2011understanding}.

During the molecular fragmentation and recombination process to generate new molecules, atoms at the cleavage sites of the fragments are the only ones that undergo significant changes, while the remaining atoms maintain relatively stable local atomic environments. To preserve these environments, simple constraints are imposed on the recombination based on the cleavage bond, which is defined by the atom types and bond orders at the break site (e.g., \ce{C-O}, \ce{N=C}). 
For carbon-started bonds, such as \ce{C-O}, only fragments with cleavage bonds starting with the terminal atom (e.g., \ce{\textbf{O}-C}, \ce{\textbf{O}-N}) are allowed. In contrast, for non-carbon-based bonds, like \ce{N=C}, pairing is allowed with any fragment that shares the same bond order (e.g., single, double, or triple). These constraints ensure that the circular neighborhood of radius 1 for all carbon atoms remains unchanged, preserving the circular neighborhood of radius 2 for most hydrogen atoms. A complete list of permissible cleavage bond pairings is provided in Supplementary Table~\ref{tab:cut_types}.

Within this framework, the NMR spectrum of a newly formed molecule $M$, assembled from fragments $F_1$ and $F_2$, is estimated by inheriting the chemical shifts of corresponding atoms from their parent molecules:
\begin{equation}
    \mathcal{X}_M = \mathcal{X}_{F_1} \cup \mathcal{X}_{F_2},
\end{equation}
and the associated encoding vectors are combined additively:
\begin{equation}
    \mathbf{v}_M = \mathbf{v}_{F_1} + \mathbf{v}_{F_2}.
\end{equation}

Candidate selection is performed in three stages to achieve high efficiency and accuracy. First, fragment pairs are pre-screened using a fast vector-based retrieval: for each fragment $F$, the top-$k$ complementary fragments $F'$ are retrieved by minimizing $\|\mathbf{v}_{\text{target}} - \mathbf{v}_F - \mathbf{v}_{F'}\|_2$ using an $\mathrm{L}^2$ distance-optimized index. Second, all candidate pairs are aggregated and re-ranked by vector similarity. Third, the top candidates are refined using set similarity between $\mathcal{X}_{F_1} \cup \mathcal{X}_{F_2}$ and the target spectrum as given in Eq.~\ref{eq:bipartite}.

Newly generated molecules undergo precise NMR prediction using the forward model (Section~\ref{subsec:method-simulation}) for final scoring. This two-tiered NMR estimation—fast inherited prediction for screening and precise forward prediction for validation—enables both scalability and high fidelity. High-similarity candidates replace lower-ranking molecules in the pool, which evolves iteratively until convergence or the maximum iteration count is reached.

\subsection{Additional constraints and prior knowledge integration}

To address various experimental constraints and incorporate prior knowledge, the algorithm features several adaptive capabilities. When the permissible elemental composition is known, fragments containing excluded elements are filtered out at each iteration, which helps reduce computational overhead. If a molecular formula is specified, candidate molecules are then evaluated against this constraint during the final stage.

When reactants are known, they are directly integrated into the initial molecule pool. This is beneficial as reactants often share structural motifs with the target molecule, thereby enhancing both the efficiency and accuracy of the search. Similarly, other types of prior knowledge—such as expert-proposed scaffolds or domain-specific molecules—can be seamlessly incorporated by including them in the initial pool.

\section*{Data and code availability}

The PubChem dataset~\cite{kim2025pubchem}, used to construct SimNMR-PubChem Database, is publicly available at \url{https://pubchem.ncbi.nlm.nih.gov}. The processed dataset and database index of the SimNMR-PubChem Database are available on Hugging Face at \url{https://huggingface.co/datasets/yqj01/SimNMR-PubChem}. 

The source code for NMR-Solver is publicly available at \url{https://github.com/YongqiJin/NMR-Solver} under the open-source MIT License. The trained model weights for NMRNet and the evaluation datasets used in this study are archived in Zenodo with DOI: \url{https://doi.org/10.5281/zenodo.16952024}.

\newpage

\bibliography{sn-bibliography}

\newpage

\begin{center}
    \Large{Supplementary Information}
\end{center}

\appendix

\section*{Supplementary Notes}
\addcontentsline{toc}{section}{Supplementary Notes}

\suppnote{Web App}

The web-based application, accessible at \url{https://www.bohrium.com/apps/nmr-toolbox}, provides user-friendly access to key functionalities of the NMR-Solver framework. The platform offers three core capabilities:

\begin{itemize}
    \item \textbf{NMR database search}: 
    The application performs spectrum-to-structure search by accepting a set of experimental $^1$H and $^{13}$C chemical shifts as input. It queries the built-in SimNMR-PubChem Database and returns a ranked list of candidate molecules, with support for filtering by element types.

    \item \textbf{Structure elucidation from NMR}: 
    The application performs automated structure elucidation by integrating the NMR-Solver framework. Given experimental $^1$H and $^{13}$C NMR spectra, it generates a diverse set of plausible candidate molecules and ranks them based on spectral similarity and prediction confidence. Allowed element types can be specified to constrain the search space, and candidate molecules can be provided as input to guide the structure elucidation process.

    \item \textbf{Chemical shift prediction and spectral matching}: 
    The application predicts $^1$H and $^{13}$C chemical shifts for given molecular structures using a deep learning model. It computes spectral similarity scores between the predicted and reference spectra, supporting structure validation and assignment.
\end{itemize}

The web interface is designed to democratize access to advanced NMR analysis, seamlessly integrating computational methods into practical chemistry workflows. It enables chemists to perform structure elucidation and spectral validation through a browser-based platform, eliminating the need for local computational infrastructure or specialized expertise.

\suppnote{Evaluation Metrics}
\label{sec:eval_metrics}

To quantitatively evaluate the performance of molecular structure prediction from NMR spectra, two complementary metrics are employed: \textbf{Recall} and \textbf{Tanimoto similarity}.

\textbf{Recall@K} measures the proportion of test cases where the correct molecular structure appears among the top-$K$ predicted candidates. Specifically, for a given sample, success is defined as the ground-truth molecular structure being ranked within the top-$K$ predictions. Recall@K is computed as the average success rate across all test samples:
\begin{equation}
    \text{Recall@K} = \frac{1}{N} \sum_{i=1}^{N} \mathbb{I} \left( \text{target}_i \in \text{top-}K\text{ predictions}_i \right),
\end{equation}
where $ \mathbb{I}(\cdot) $ is the indicator function that returns 1 when the condition is satisfied and 0 otherwise.

In addition to exact structure matching, \textbf{Tanimoto similarity}~\cite{tanimoto1958elementary, bajusz2015tanimoto} is used to measure the molecular similarity between the ground-truth and predicted structures based on their Morgan fingerprints~\cite{morgan1965generation, rogers2010extended}. 
In this work, Morgan fingerprints are computed using the RDKit toolkit as 2048-bit vectors with a radius of $r=2$, capturing the local structural features of each molecule. 
The Tanimoto coefficient is defined as:
\begin{equation}
    \text{Tanimoto}(A, B) = \frac{|A \cap B|}{|A \cup B|},
\end{equation}
where $ A $ and $ B $ are the fingerprint bit vectors of the two molecules. A higher Tanimoto score indicates greater structural similarity.

We also define \textbf{Tanimoto@K} to measure the maximum Tanimoto similarity between the ground-truth molecule and any of the top-$K$ predicted structures. This metric provides insight into how structurally similar the best-ranked prediction is to the true molecule, even if an exact match is not achieved. Formally:
\begin{equation}
    \text{Tanimoto@K} = \frac{1}{N} \sum_{i=1}^{N} \max_{j=1,\dots,K} 
    \Big( \text{Tanimoto}\big(f_{\rm Morgan}(\text{target}_i), f_{\rm Morgan}(\text{prediction}_{i,j})\big) \Big)
\end{equation}

Both metrics are reported across multiple values of $ K $ (e.g., $ K = 1, 3, 10 $) to provide a comprehensive view of the model’s performance in terms of both exact structure retrieval and structural closeness to the ground truth.

\suppnote{Ablation Study}

To independently evaluate the contribution of the optimization stage, an ablation study is conducted by removing target molecules from the initial retrieval results, forming a contrast with the default process. This ensures that any remaining performance improvements are attributed to the optimization process refining non-trivial candidates, rather than retrieval recall alone.

The experimental results, shown in Supplementary Table~\ref{tab:result_ablation}, demonstrate that the method exhibits only a marginal performance drop (less than 2\% across all settings for both $^1$H and $^{13}$C NMR). This indicates that the optimization module retains its effectiveness even when handling molecules not present in the database. Such a small drop suggests that the optimization process is the primary driver of the improvements, rather than relying heavily on retrieval recall.

Through this ablation study, the robustness of the optimization process is underscored. Furthermore, the results show that NMR-Solver is capable of performing NMR analysis for molecules previously unknown to humans, further demonstrating its generalization ability.

\suppnote{Incorporation of Multiplicity Patterns in the Set Similarity}
\label{sec:multiplet-type}

For experimental $^1$H NMR spectra, multiplicity patterns are typically reported in the spectral data and are directly extracted. For simulated spectra, multiplicity patterns are estimated based on the first-order coupling effects, which predicts the splitting pattern from the number of neighboring protons in the molecular structure. 

The multiplicity patterns are incorporated into the matching function as multiplicative weights, \( \omega_{ij} \), which modifies the original scoring function. This additional information helps to refine the peak assignments and improve the overall similarity score.

Details of the exact implementation are as follows:

\begin{equation} \label{eq:multiplet-modify}
    S(\mathcal{X}, \mathcal{Y}) = \frac{1}{\sqrt{mn}} \max_{P \in \mathcal{P}} \sum_{(i,j) \in P} \omega_{ij} f(x_i, y_j),
\end{equation}

where \( \omega_{ij} \) is set to \( w_1 \) (default value: 1.0) if the multiplicity pattern of \( x_i \) matches that of \( y_j \), and to \( w_2 \) (default value: 0.8) otherwise.

\suppnote{Comparison of Spectral Similarity Metrics} \label{sec:similarity}

Within the NMR-Solver framework, we employ two complementary similarity scoring methods: \textbf{vector similarity} and \textbf{set similarity}. To demonstrate their effectiveness and robustness in capturing structural and spectral correspondence, these metrics are compared against conventional approaches such as the \textbf{Wasserstein distance}~\cite{rubner2000earth}, which has been widely used for spectrum matching in chemical informatics.

The performance of both vector similarity and set similarity is evaluated against the Wasserstein distance on a test set of 1000 randomly selected molecules from our NMR database. For each query molecule, a set of 1000 candidate molecules is retrieved using vector index. The goal is to identify the correct molecule from this set of candidates.
To simulate realistic experimental variations, Gaussian-distributed chemical shift perturbations were applied to the query spectra, along with random peak deletions and insertions to model signal overlap and impurity effects.

The results (Supplementary Figure~\ref{fig:result_similarity}) demonstrate that the set similarity measure exhibits strong robustness against both spectral shift noise and peak deletion/insertion perturbations, significantly outperforming the alternative methods under realistic degradation conditions. In contrast, the vector similarity method degrades rapidly with increasing noise levels (Supplementary Figure~\ref{fig:result_similarity-a}), reflecting its poor resilience to large deviations between predicted and experimental chemical shifts. The Wasserstein distance performs well under shift noise but deteriorates noticeably as the probability of random peak deletion or insertion increases (Supplementary Figure~\ref{fig:result_similarity-b}), highlighting its vulnerability to peak overlap and spurious peaks. Detailed experimental results are provided in Supplementary Tables~\ref{tab:similarity-1} and~\ref{tab:similarity-2}.

This comparative analysis underscores the practical advantage of the set similarity as a scoring function in real-world applications, where spectral data are often incomplete or noisy. Moreover, it enables the robust identification of molecules with partial spectral overlap, supporting queries for compounds that share chemically meaningful fragments—such as functional groups or ring systems—even in the absence of full structural similarity.

\clearpage

\section*{Supplementary Tables}
\addcontentsline{toc}{section}{Supplementary Tables}

\begin{supptable}[h]
    \supptablecaption{Hyperparameters for similarity metrics}{}
    \label{tab:scoring_parameters}
    \begin{tabular*}{\textwidth}{@{\extracolsep{\fill}} l c c }
    \toprule
        Parameter & $^1$H NMR & $^{13}$C NMR \\
    \midrule
        \textbf{Vector Similarity} & & \\
        $\sigma$ (Guassian kernel) & 0.3 & 2 \\
        range & $[-1, 15]$ & $[-10, 230]$ \\
        dimension & 128 & 128 \\
    \midrule
        \textbf{Set Similarity} & & \\
        $\sigma$ (Guassian kernel) & 1 & 10 \\
    \bottomrule
    \end{tabular*}
\end{supptable}

\begin{supptable}[h]
    \supptablecaption{Hyperparameters for database index}{}
    \begin{tabular*}{\textwidth}{@{\extracolsep{\fill}} cccc }
    \toprule
        D & HNSW: M & HNSW: efConstruction & HNSW: efSearch \\
    \midrule
        256 & 32 & 600 & 2000 \\
    \bottomrule
    \end{tabular*}
    \label{tab:database_parameters}
\end{supptable}

\begin{supptable}[h]
    \supptablecaption{Hyperparameters for FB-MO}{}
    \begin{tabular*}{\textwidth}{@{\extracolsep{\fill}} lll }
    \toprule
        Hyperparameter & Value & Description \\
    \midrule
        num\_search     & 1000      & Number of candidate molecules to explore during search \\
        num\_pool       & 1000      & Size of the candidate pool \\
        num\_filter\_pair & 200000  & Max number of molecular pairs to consider during filtering \\
        num\_filter\_mol & 1000     & Max number of molecules to retain after filtering \\
    \bottomrule
    \end{tabular*}
    \label{tab:fbmo_parameters}
\end{supptable}

\begin{supptable}[h]
    \supptablecaption{Comparison with current methods on the experimental dataset (stereochemistry ignored)}{}
    \label{tab:result_exp_wo_stereo}
    \begin{tabular*}{\textwidth}{@{\extracolsep{\fill}} l l c c c c c c }
    \toprule
        NMR Type & Condition & Top-1(\%) & Top-3(\%) & Top-10(\%) & Tani\footnotemark[1]@1 & Tani@3 & Tani@10\\
    \midrule
        \multicolumn{8}{l}{\textbf{NMR-to-Structure~\cite{alberts2024unraveling}}\footnotemark[2]} \\
    \midrule
        \multirow{3}{*}{$^1$H}
        & No & 0.67 & 0.89 & 1.56 & 0.222 & 0.249 & 0.282 \\
        & Elements\footnotemark[3] & — & — & — & — & — & — \\
        & Formula\footnotemark[4] & 7.78 & 10.67 & 12.22 & 0.314 & 0.370 & 0.408 \\
    \midrule
        \multirow{3}{*}{$^{13}$C}
        & No & 2.89 & 6.00 & 8.00 & 0.252 & 0.307 & 0.353 \\
        & Elements & — & — & — & — & — & — \\
        & Formula & 12.89 & 17.33 & 20.22 & 0.366 & 0.418 & 0.470 \\
        \midrule
        \multirow{3}{*}{$^1$H + $^{13}$C}
        & No & 4.89 & 7.78 & 11.56 & 0.325 & 0.376 & 0.428 \\
        & Elements & — & — & — & — & — & — \\
        & Formula & 14.44 & 20.22 & 21.78 & 0.404 & 0.468 & 0.518 \\
    \midrule
        \multicolumn{8}{l}{\textbf{NMR-Solver (Ours)}} \\
    \midrule
        \multirow{3}{*}{$^1$H}
        & No & 0.67 & 1.11 & 3.78 & 0.196 & 0.244 & 0.294 \\
        & Elements & 3.11 & 7.33 & 12.67 & 0.269 & 0.342 & 0.410 \\
        & Formula & 20.00 & 27.33 & 31.78 & 0.382 & 0.445 & 0.488 \\
    \midrule
        \multirow{3}{*}{$^{13}$C}
        & No & 12.22 & 17.33 & 23.78 & 0.371 & 0.436 & 0.499 \\
        & Elements & 24.00 & 29.56 & 34.67 & 0.488 & 0.540 & 0.603 \\
        & Formula & 45.11 & 50.67 & 54.89 & 0.591 & 0.631 & 0.659 \\
    \midrule
        \multirow{3}{*}{$^1$H + $^{13}$C}
        & No & 25.56 & 32.89 & 37.33 & 0.524 & 0.594 & 0.643 \\
        & Elements & 38.44 & 45.11 & 51.78 & 0.624 & 0.684 & 0.736 \\
        & Formula & \textbf{52.89} & \textbf{61.33} & \textbf{67.33} & \textbf{0.709} & \textbf{0.766} & \textbf{0.796} \\
    \bottomrule
    \end{tabular*}
    \footnotetext[1]{Tanimoto similarity.}
    \footnotetext[2]{The models were reimplemented and trained using the code and simulated data provided by the original authors.}
    \footnotetext[3]{The method does not support using elemental composition as an input condition.}
    \footnotetext[4]{The molecular formula, in addition to being used as input for the models, was also consistently applied as a filtering criterion for the final candidate structures across all methods.}
\end{supptable}

\begin{supptable}[h]
    \supptablecaption{Comparison with current methods on the experimental dataset (stereochemistry preserved)}{}
    \label{tab:result_exp_with_stereo}
    \begin{tabular*}{\textwidth}{@{\extracolsep{\fill}} l l c c c c c c }
    \toprule
        NMR Type & Condition & Top-1(\%) & Top-3(\%) & Top-10(\%) & Tani@1 & Tani@3 & Tani@10\\
    \midrule
        \multicolumn{8}{l}{\textbf{NMR-to-Structure}~\cite{alberts2024unraveling}} \\
    \midrule
        \multirow{3}{*}{$^1$H}
        & No & 0.67 & 0.89 & 1.11 & 0.219 & 0.246 & 0.276 \\
        & Elements & — & — & — & — & — & — \\
        & Formula & 6.44 & 8.67 & 9.78 & 0.307 & 0.364 & 0.400 \\
    \midrule
        \multirow{3}{*}{$^{13}$C}
        & No & 2.89 & 5.33 & 6.44 & 0.250 & 0.303 & 0.347 \\
        & Elements & — & — & — & — & — & — \\
        & Formula & 9.55 & 14.00 & 16.44 & 0.360 & 0.410 & 0.460 \\
        \midrule
        \multirow{3}{*}{$^1$H + $^{13}$C}
        & No & 4.00 & 7.11 & 10.00 & 0.320 & 0.371 & 0.419 \\
        & Elements & — & — & — & — & — & — \\
        & Formula & 10.88 & 16.44 & 17.33 & 0.391 & 0.455 & 0.500 \\
    \midrule
        \multicolumn{8}{l}{\textbf{NMR-Solver (Ours)}} \\
    \midrule
        \multirow{3}{*}{$^1$H}
        & No & 0.00 & 1.33 & 2.44 & 0.196 & 0.243 & 0.287 \\
        & Elements & 2.00 & 5.78 & 9.78 & 0.266 & 0.339 & 0.403 \\
        & Formula & 12.00 & 21.78 & 27.33 & 0.350 & 0.414 & 0.456 \\
    \midrule
        \multirow{3}{*}{$^{13}$C}
        & No & 8.89 & 13.33 & 19.11 & 0.364 & 0.426 & 0.488 \\
        & Elements & 15.78 & 24.44 & 30.67 & 0.462 & 0.531 & 0.592 \\
        & Formula & 30.0 & 42.22 & 45.33 & 0.563 & 0.605 & 0.628 \\
    \midrule
        \multirow{3}{*}{$^1$H + $^{13}$C}
        & No & 15.11 & 25.56 & 29.11 & 0.503 & 0.573 & 0.618 \\
        & Elements & 22.44 & 34.89 & 40.89 & 0.587 & 0.656 & 0.702 \\
        & Formula & \textbf{31.56} & \textbf{47.56} & \textbf{53.78} & \textbf{0.659} & \textbf{0.720} & \textbf{0.758} \\
    \bottomrule
    \end{tabular*}
\end{supptable}

\begin{supptable}[h]
    \supptablecaption{Comparison of recall rates for NMR-Solver with and without reactants}{\footnotemark[1]}
    \label{tab:result_reaction}
    \begin{tabular*}{\textwidth}{@{\extracolsep{\fill}} l l c c c c c c }
    \toprule
        NMR Type & Condition & Top-1(\%) & Top-3(\%) & Top-10(\%) & Tani@1 & Tani@3 & Tani@10\\
    \midrule
        \multicolumn{8}{l}{\textbf{NMR-Solver (without reactants)}} \\
    \midrule
        \multirow{3}{*}{$^1$H}
        & No & 0.67 & 1.11 & 3.78 & 0.196 & 0.244 & 0.294 \\
        & Elements & 3.11 & 7.33 & 12.67 & 0.269 & 0.342 & 0.410 \\
        & Formula & 20.00 & 27.33 & 31.78 & 0.382 & 0.445 & 0.488 \\
    \midrule
        \multirow{3}{*}{$^{13}$C}
        & No & 12.22 & 17.33 & 23.78 & 0.371 & 0.436 & 0.499 \\
        & Elements & 24.00 & 29.56 & 34.67 & 0.488 & 0.540 & 0.603 \\
        & Formula & 45.11 & 50.67 & 54.89 & 0.591 & 0.631 & 0.659 \\
    \midrule
        \multirow{3}{*}{$^1$H + $^{13}$C}
        & No & 25.56 & 32.89 & 37.33 & 0.524 & 0.594 & 0.643 \\
        & Elements & 38.44 & 45.11 & 51.78 & 0.624 & 0.684 & 0.736 \\
        & Formula & 52.89 & 61.33 & 67.33 & 0.709 & 0.766 & 0.796 \\
    \midrule
        \multicolumn{8}{l}{\textbf{NMR-Solver (with reactants)}} \\
    \midrule
        \multirow{3}{*}{$^1$H}
        & No\footnotemark[2] & — & — & — & — & — & — \\
        & Elements & 3.33 & 8.00 & 15.11 & 0.285 & 0.368 & 0.441 \\
        & Formula  & 25.11 & 34.22 & 40.44 & 0.433 & 0.507 & 0.556 \\
    \midrule
        \multirow{3}{*}{$^{13}$C}
        & No & — & — & — & — & — & — \\
        & Elements & 26.44 & 33.11 & 38.67 & 0.518 & 0.583 & 0.649 \\
        & Formula  & 50.89 & 56.89 & 61.56 & 0.647 & 0.687 & 0.718 \\
    \midrule
        \multirow{3}{*}{$^1$H + $^{13}$C}
        & No & — & — & — & — & — & — \\
        & Elements & 42.89 & 50.67 & 57.78 & 0.664 & 0.726 & 0.779 \\
        & Formula  & \textbf{60.22} & \textbf{69.11} & \textbf{76.22} & \textbf{0.767} & \textbf{0.824} & \textbf{0.858} \\
    \bottomrule
    \end{tabular*}
    \footnotetext[1]{Stereochemistry not considered.}
    \footnotetext[2]{When reactants are provided, the permissible elemental composition of the products is often constrained.}
\end{supptable}

\begin{supptable}[h]
    \supptablecaption{Ablation study on the contribution of the optimization stage}{\footnotemark[1]}
    \label{tab:result_ablation}
    \begin{tabular*}{\textwidth}{@{\extracolsep{\fill}} l l c c c c c c }
    \toprule
        NMR Type & Condition & Top-1(\%) & Top-3(\%) & Top-10(\%) & Tani@1 & Tani@3 & Tani@10\\
    \midrule
        \multicolumn{8}{l}{\textbf{NMR-Solver (default)}} \\
    \midrule
        \multirow{3}{*}{$^1$H}
        & No & 0.67 & 1.11 & 3.78 & 0.196 & 0.244 & 0.294 \\
        & Elements & 3.11 & 7.33 & 12.67 & 0.269 & 0.342 & 0.410 \\
        & Formula  & 20.00 & 27.33 & 31.78 & 0.382 & 0.445 & 0.488 \\
    \midrule
        \multirow{3}{*}{$^{13}$C}
        & No & 12.22 & 17.33 & 23.78 & 0.371 & 0.436 & 0.499 \\
        & Elements & 24.00 & 29.56 & 34.67 & 0.488 & 0.540 & 0.603 \\
        & Formula  & 45.11 & 50.67 & 54.89 & 0.591 & 0.631 & 0.659 \\
    \midrule
        \multirow{3}{*}{$^1$H + $^{13}$C}
        & No & 25.56 & 32.89 & 37.33 & 0.524 & 0.594 & 0.643 \\
        & Elements & 38.44 & 45.11 & 51.78 & 0.624 & 0.684 & 0.736 \\
        & Formula  & \textbf{52.89} & \textbf{61.33} & \textbf{67.33} & \textbf{0.709} & \textbf{0.766} & \textbf{0.796} \\
    \midrule
        \multicolumn{8}{l}{\textbf{NMR-Solver (removing target from candidates)}} \\
    \midrule
        \multirow{6}{*}{$^1$H}
        & No 
            & 0.22 & 0.89 & 2.89 & 0.192 & 0.241 & 0.286 \\
            & & (-0.45) & (-0.22) & (-0.89) & (-0.004) & (-0.003) & (-0.008) \\
        & Elements 
            & 2.00 & 6.00 & 10.22 & 0.260 & 0.329 & 0.392 \\
            & & (-1.11) & (-1.33) & (-2.45) & (-0.009) & (-0.013) & (-0.018) \\
        & Formula  
            & 16.67 & 23.33 & 28.00 & 0.359 & 0.423 & 0.469 \\
            & & (-3.33) & (-4.00) & (-3.78) & (-0.023) & (-0.022) & (-0.019) \\
    \midrule
        \multirow{6}{*}{$^{13}$C}
        & No 
            & 12.22 & 16.89 & 22.22 & 0.369 & 0.432 & 0.486 \\
            & & (-0.00) & (-0.44) & (-1.56) & (-0.002) & (-0.004) & (-0.013) \\
        & Elements 
            & 22.00 & 28.22 & 34.00 & 0.473 & 0.532 & 0.596 \\
            & & (-2.00) & (-1.34) & (-0.67) & (-0.015) & (-0.008) & (-0.007) \\
        & Formula  
            & 41.56 & 46.89 & 50.67 & 0.567 & 0.608 & 0.634 \\
            & & (-3.55) & (-3.78) & (-4.22) & (-0.024) & (-0.023) & (-0.025) \\
    \midrule
        \multirow{6}{*}{$^1$H + $^{13}$C}
        & No 
            & 24.60 & 32.80 & 36.67 & 0.517 & 0.592 & 0.637 \\
            & & (-0.96) & (-0.09) & (-0.66) & (-0.007) & (-0.002) & (-0.006) \\
        & Elements 
            & 38.22 & 43.56 & 50.00 & 0.623 & 0.674 & 0.726 \\
            & & (-0.22) & (-1.55) & (-1.78) & (-0.001) & (-0.010) & (-0.010) \\
        & Formula  
            & \textbf{51.33} & \textbf{60.00} & \textbf{65.56} & \textbf{0.700} & \textbf{0.758} & \textbf{0.786} \\
            & & \textbf{(-1.56)} & \textbf{(-1.33)} & \textbf{(-1.77)} & \textbf{(-0.009)} & \textbf{(-0.008)} & \textbf{(-0.010)} \\
    \bottomrule
    \end{tabular*}
    \footnotetext[1]{Stereochemistry not considered.}
\end{supptable}

\begin{supptable}
    \supptablecaption{\texorpdfstring{Performance of spectral similarity metrics under varying noise and perturbation conditions ($p=0$)}{Performance of spectral similarity metrics under varying noise and perturbation conditions (p=0)}}{
    The parameter $\sigma$ indicates the magnitude of spectral noise, modeled as Gaussian deviations in chemical shifts, and $p$ denotes the probability of random peak insertion or deletion, simulating signal overlap and impurity effects. Cells shaded in \colorbox{gray!20}{gray} indicate relatively noticeable poorer performance (more than 5\% lower than the best performance in the row).
    }

    \label{tab:similarity-1}
\begin{tabular*}{\textwidth}{@{\extracolsep{\fill}} l c c c c c }
    \toprule
    NMR Type & $p$ & $\sigma$ & Wasserstein & Vector & Set \\
    \midrule
    \multicolumn{6}{c}{\textit{Top-1 Recall (\%)}} \\
    \midrule
    \multirow{4}{*}{$^1$H} 
        & \multirow{4}{*}{0.0} & 0.0 & 100 & 100 & 100 \\
        &                       & 0.1 & 75.2 & \cellcolor{gray!20}55.8 & 76.9 \\
        &                       & 0.2 & 27.4 & \cellcolor{gray!20}7.4 & 31.1 \\
        &                       & 0.3 & 5.7 & \cellcolor{gray!20}0.9 & 7.9 \\
    \midrule
    \multirow{4}{*}{$^{13}$C} 
        & \multirow{4}{*}{0.0} & 0.0 & 100 & 100 & 100 \\
        &                       & 0.1 & 86.3 & \cellcolor{gray!20}80.2 & 87.9 \\
        &                       & 0.2 & 47.6 & \cellcolor{gray!20}16.8 & 50.9 \\
        &                       & 0.3 & 16.0 & \cellcolor{gray!20}2.4 & 17.4 \\
    \midrule
    \multirow{4}{*}{$^1$H + $^{13}$C} 
        & \multirow{4}{*}{0.0} & 0.0 & 100 & 100 & 100 \\
        &                       & 0.1 & 96.7 & 93.3 & 95.5 \\
        &                       & 0.2 & 85.2 & \cellcolor{gray!20}52.9 & 86.2 \\
        &                       & 0.3 & 56.6 & \cellcolor{gray!20}13.5 & 57.7 \\
    \midrule
    \multicolumn{6}{c}{\textit{Top-10 Recall (\%)}} \\
    \midrule
    \multirow{4}{*}{$^1$H} 
        & \multirow{4}{*}{0.0} & 0.0 & 100 & 100 & 100 \\
        &                       & 0.1 & 91.9 & \cellcolor{gray!20}81.3 & 92.8 \\
        &                       & 0.2 & 50.9 & \cellcolor{gray!20}20.4 & 53.0 \\
        &                       & 0.3 & 13.2 & \cellcolor{gray!20}3.0 & 15.4 \\
    \midrule
    \multirow{4}{*}{$^{13}$C} 
        & \multirow{4}{*}{0.0} & 0.0 & 100 & 100 & 100 \\
        &                       & 0.1 & 99.2 & 96.5 & 99.0 \\
        &                       & 0.2 & 70.1 & \cellcolor{gray!20}38.3 & 70.7 \\
        &                       & 0.3 & 28.3 & \cellcolor{gray!20}7.3 & 28.4 \\
    \midrule
    \multirow{4}{*}{$^1$H + $^{13}$C} 
        & \multirow{4}{*}{0.0} & 0.0 & 100 & 100 & 100 \\
        &                       & 0.1 & 100 & 99.5 & 99.9 \\
        &                       & 0.2 & 97.0 & \cellcolor{gray!20}80.0 & 95.8 \\
        &                       & 0.3 & 71.0 & \cellcolor{gray!20}30.8 & 68.5 \\
    \bottomrule

\end{tabular*}
\end{supptable}

\begin{supptable}
    \supptablecaption{\texorpdfstring{Performance of spectral similarity metrics under varying noise and perturbation conditions ($p=0.2$)}{Performance of spectral similarity metrics under varying noise and perturbation conditions (p=0.2)}}{
    The parameter $\sigma$ indicates the magnitude of spectral noise, modeled as Gaussian deviations in chemical shifts, and $p$ denotes the probability of random peak insertion or deletion, simulating signal overlap and impurity effects. Cells shaded in \colorbox{gray!20}{gray} indicate relatively noticeable poorer performance (more than 5\% lower than the best performance in the row).
    }

    \label{tab:similarity-2}
\begin{tabular*}{\textwidth}{@{\extracolsep{\fill}} l c c c c c }
    \toprule
    NMR Type & $p$ & $\sigma$ & Wasserstein & Vector & Set \\
    \midrule
    \multicolumn{6}{c}{\textit{Top-1 Recall (\%)}} \\
    \midrule
    \multirow{4}{*}{$^1$H} 
        & \multirow{4}{*}{0.2} & 0.0 & \cellcolor{gray!20}88.3 & 95.5 & \cellcolor{gray!20}88.3 \\
        &                       & 0.1 & 63.0 & \cellcolor{gray!20}52.9 & 67.7 \\
        &                       & 0.2 & 22.5 & \cellcolor{gray!20}7.0  & 27.3 \\
        &                       & 0.3 & 4.6  & \cellcolor{gray!20}0.8  & 7.0 \\
    \midrule
    \multirow{4}{*}{$^{13}$C} 
        & \multirow{4}{*}{0.2} & 0.0 & \cellcolor{gray!20}83.9 & 99.1 & \cellcolor{gray!20}90.7 \\
        &                       & 0.1 & \cellcolor{gray!20}70.9 & 77.1 & 79.7 \\
        &                       & 0.2 & \cellcolor{gray!20}39.0 & \cellcolor{gray!20}16.3 & 45.9 \\
        &                       & 0.3 & 13.2 & \cellcolor{gray!20}2.3  & 15.4 \\
    \midrule
    \multirow{4}{*}{$^1$H + $^{13}$C} 
        & \multirow{4}{*}{0.2} & 0.0 & \cellcolor{gray!20}90.3 & 100 & 98.3 \\
        &                       & 0.1 & \cellcolor{gray!20}84.0 & 92.6 & 93.3 \\
        &                       & 0.2 & \cellcolor{gray!20}72.2 & \cellcolor{gray!20}51.9 & 83.3 \\
        &                       & 0.3 & \cellcolor{gray!20}47.6 & \cellcolor{gray!20}12.9 & 55.0 \\
    \midrule
    \multicolumn{6}{c}{\textit{Top-10 Recall (\%)}} \\
    \midrule
    \multirow{4}{*}{$^1$H} 
        & \multirow{4}{*}{0.2} & 0.0 & \cellcolor{gray!20}91.6 & 98.4 & 94.0 \\
        &                       & 0.1 & \cellcolor{gray!20}80.0 & \cellcolor{gray!20}78.0 & 86.3 \\
        &                       & 0.2 & \cellcolor{gray!20}42.9 & \cellcolor{gray!20}19.3 & 49.0 \\
        &                       & 0.3 & 11.0 & \cellcolor{gray!20}2.7  & 14.1 \\
    \midrule
    \multirow{4}{*}{$^{13}$C} 
        & \multirow{4}{*}{0.2} & 0.0 & \cellcolor{gray!20}87.8 & 100 & 97.5 \\
        &                       & 0.1 & \cellcolor{gray!20}85.5 & 95.0 & 96.0 \\
        &                       & 0.2 & \cellcolor{gray!20}59.3 & \cellcolor{gray!20}36.8 & 67.5 \\
        &                       & 0.3 & 24.1 & \cellcolor{gray!20}7.0  & 26.5 \\
    \midrule
    \multirow{4}{*}{$^1$H + $^{13}$C} 
        & \multirow{4}{*}{0.2} & 0.0 & 95.4 & 100 & 99.9 \\
        &                       & 0.1 & \cellcolor{gray!20}93.0 & 99.5  & 99.7 \\
        &                       & 0.2 & \cellcolor{gray!20}87.9 & \cellcolor{gray!20}78.5  & 95.2 \\
        &                       & 0.3 & 63.1 & \cellcolor{gray!20}30.1  & 66.9 \\
    \bottomrule
\end{tabular*}

\end{supptable}

\begin{supptable}
    \supptablecaption{Allowed cleavage bond pairings in FB-MO}{For each query fragment cleavage bond, the table lists valid complementary cleavage bonds that satisfy the local environment preservation constraint. Here, \textbf{C} denotes a carbon atom, and \textbf{D} denotes a non-carbon atom (e.g., N, O, S). \textbf{A}$_1$, \textbf{A}$_2$, and \textbf{A}$_3$ represent any chemical atom. Halogens (F, Cl, Br, I) are treated as a single element class for pairing purposes, reflecting their similar bonding behavior in fragment recombination.}
    \label{tab:cut_types}
    \begin{tabular*}{\textwidth}{@{\extracolsep{\fill}} c c }
        \toprule
        Query Cleavage Bond & Allowed Complementary Cleavage Bonds \\
        \midrule
        \ce{C-A1}    & \ce{A1-A2} \\
        \ce{C=A1}    & \ce{A1=A2} \\
        \ce{C#A1}    & \ce{A1#A2} \\
        \ce{D-A3}    & \ce{A1-A2} \\
        \ce{D=A3}    & \ce{A1=A2} \\
        \ce{D#A3}    & \ce{A1#A2} \\
        \bottomrule
    \end{tabular*}
\end{supptable}

\clearpage

\section*{Supplementary Figures}
\addcontentsline{toc}{section}{Supplementary Figures}

\begin{suppfigure}[h]
    \begin{subfigure}[t]{0.45\textwidth}
        \includegraphics[width=\linewidth]{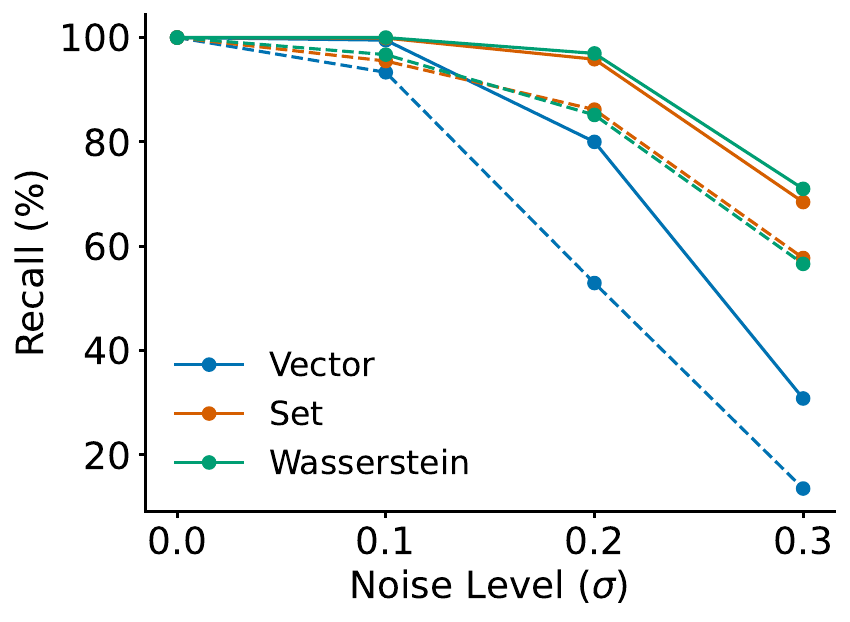}
        \put(-175,115){\textbf{a}} \phantomsubcaption
        \label{fig:result_similarity-a}
    \end{subfigure}
    \hspace{0.05\textwidth}
    \begin{subfigure}[t]{0.45\textwidth}
        \includegraphics[width=\linewidth]{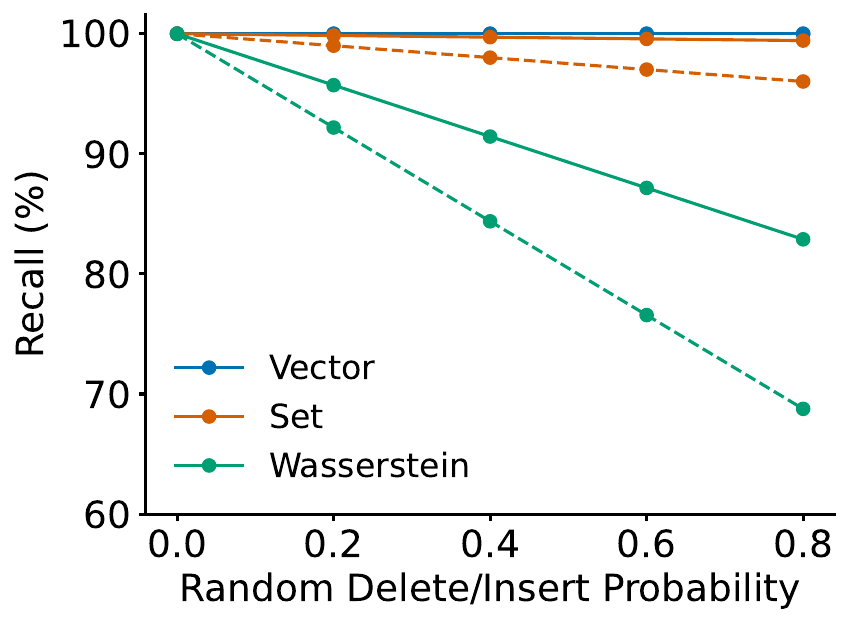}
        \put(-175,115){\textbf{b}} \phantomsubcaption
        \label{fig:result_similarity-b}
    \end{subfigure}

    \suppfigurecaption{Comparison of spectral similarity metrics under varying noise and perturbation conditions}{
    \textbf{a.} Recall vs.\ spectral noise level, where chemical shift deviations are modeled as Gaussian noise with a standard deviation of $\sigma$ ppm for $^1$H NMR and $10\sigma$ ppm for $^{13}$C NMR. 
    \textbf{b.} Recall vs.\ random peak deletion/insertion probability, simulating signal overlap and impurity effects.
    Dashed lines indicate top-1 recall, while solid lines indicate top-10 recall.
    }
    \label{fig:result_similarity}
\end{suppfigure}

\begin{suppfigure}[h]
    \includegraphics[width=1.0\textwidth]{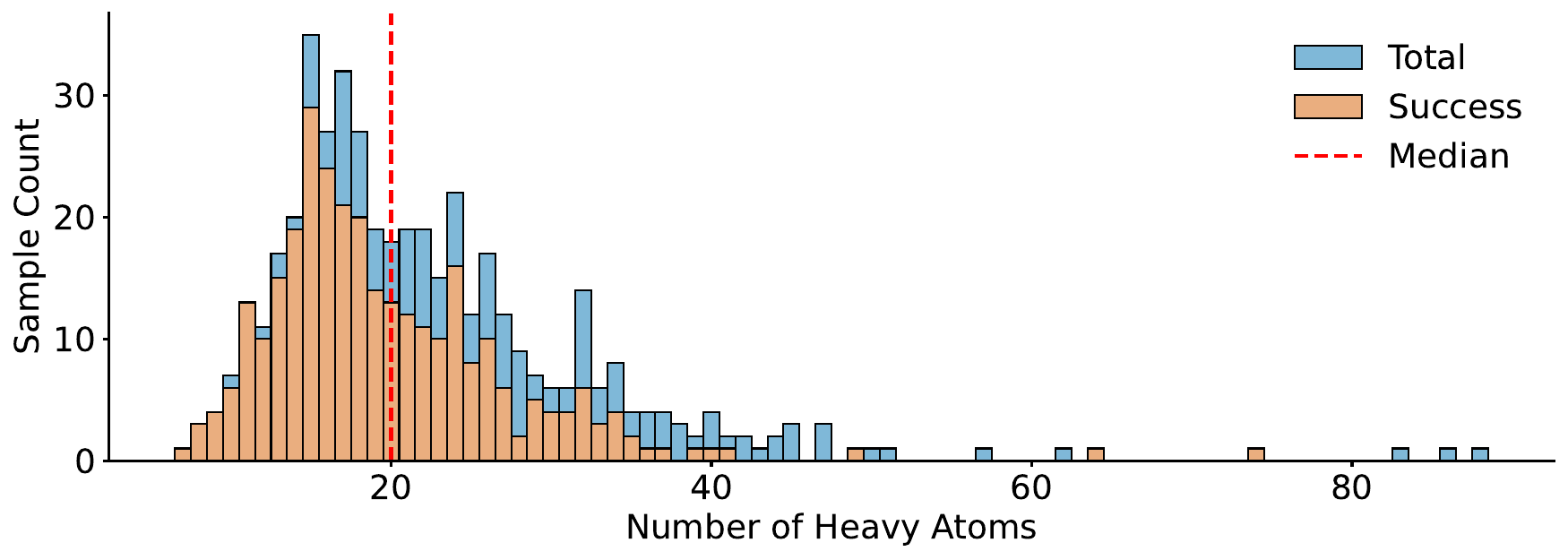}
    \suppfigurecaption{Atom count distribution of the literature dataset}{Correct prediction means the ground truth appears in the top 10 solutions.}
\end{suppfigure}

\begin{suppfigure}[h]
    \includegraphics[width=1.0\textwidth]{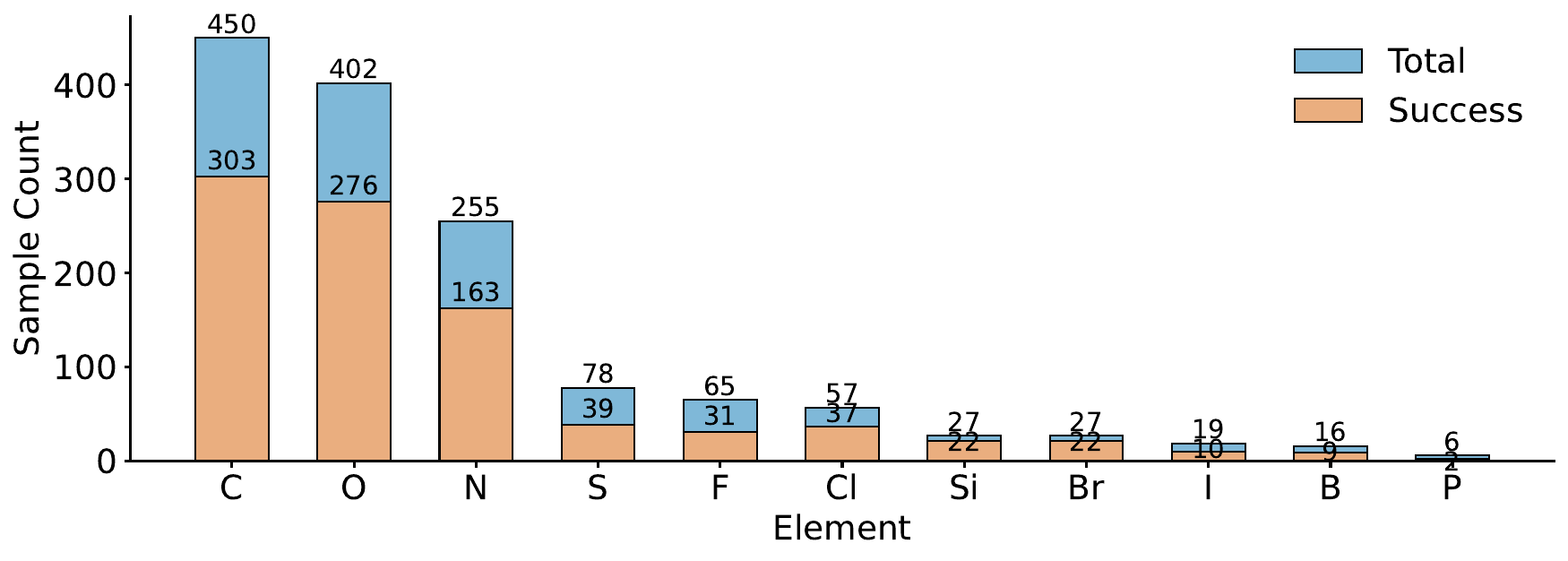}
    \suppfigurecaption{Element statistics of the literature dataset}{Correct prediction means the ground truth appears in the top 10 solutions.}
\end{suppfigure}

\end{document}